\documentclass[]{jfm}
\usepackage{natbib}
\usepackage{amsmath}
\usepackage[dvips]{graphicx}
\usepackage{psfrag}
\usepackage{tikz}
\usepackage{color}
\usepackage{mathtools}
\usepackage{lmodern}

\newcommand{\e}{\mbox{e}}
\newcommand{\p}{\partial}

\newcommand{\dif}{\mathrm{d}}
\newcommand{\turb}{\mathcal{T}}
\newcommand{\pipe}[1]{\left.#1\right|}
\newcommand{\mymk}[1]{\tikz \node[draw,circle, inner sep=0pt, minimum size=4mm]{#1};}
\newcommand{\mymktiny}[1]{\tikz \node[draw,circle, inner sep=0pt, minimum size=3mm]{#1};}
\newcommand{\plus}{{\tiny $^+$}}

\newcommand{\averxz}[1]{\widetilde{#1}}
\newcommand{\averper}[1]{\widehat{#1}}
\newcommand{\averph}[1]{\left\langle {#1} \right\rangle}
\newcommand{\averglob}[1]{\left[ {#1} \right]_g}

\begin{document}
\title[]{
Changes in turbulent dissipation \\ in a channel flow with oscillating walls}
\author[P. Ricco, C. Ottonelli, Y. Hasegawa \&  M. Quadrio]{
By P\ls I\ls E\ls R\ls R\ls E\ls\ns R\ls I\ls C\ls C\ls O$^1$, \ls
C\ls L\ls A\ls U\ls D\ls I\ls O\ls \ns O\ls T\ls T\ls O\ls N\ls E\ls L\ls L\ls I$^2$\footnote{Present address: ONERA, D\'epartement d'A\'erodynamique Fondamentale et Exp\'erimentale, 8, rue des Vertugadins, 92190 Meudon, France.},\ls \\
Y\ls O\ls S\ls U\ls K\ls E\ls \ns H\ls A\ls S\ls E\ls G\ls A\ls W\ls A\ls$^{3,4}$, \\
\and
M\ls A\ls U\ls R\ls I\ls Z\ls I\ls O\ls \ns Q\ls U\ls A\ls D\ls R\ls I\ls O$^2$ 
}
\affiliation{
$^1$Department of Mechanical Engineering, The University of Sheffield, \\ Mappin Street, Sheffield S1 3JD, United Kingdom \\
$^2$Dipartimento di Ingegneria Aerospaziale del Politecnico di Milano, \\ via La Masa 34, 20156 Milano, Italy \\
$^3$Center of Smart Interfaces, TU Darmstadt, \\ Petersenstr. 32, 64287, Darmstadt, Germany \\
$^4$Department of Mechanical Engineering, The University of Tokyo, \\ Hongo 7-3-1, Bunkyo-ku, Tokyo 113-8656, Japan\\[\affilskip]
}

\maketitle

\begin{abstract}

Harmonic oscillations of the walls of a turbulent plane channel flow are studied by direct numerical simulations to improve our understanding of the physical mechanism for skin-friction drag reduction. The simulations are carried out at constant pressure gradient in order to define an unambiguous inner scaling: in this case, drag reduction manifests itself as an increase of mass flow rate. Energy and enstrophy balances, carried out to emphasize the role of the oscillating spanwise shear layer, show that the viscous dissipations of the mean flow and of the turbulent fluctuations increase with the mass flow rate, and the relative importance of the latter decreases. We then focus on the turbulent enstrophy: through an analysis of the temporal evolution from the beginning of the wall motion, the dominant, oscillation-related term in the turbulent enstrophy is shown to cause the turbulent dissipation to be enhanced in absolute terms, before the slow drift towards the new quasi-equilibrium condition. This mechanism is found to be responsible for the increase in mass flow rate. We finally show that the time-average volume integral of the dominant term relates linearly to the drag reduction. 
\end{abstract}

\section{Introduction}

The reduction of skin-friction drag in wall-bounded turbulent flows is an important and challenging area of fluid mechanics. Its difficulty lies both in the extreme complexity of the physics underlying turbulence and in the resistance of such flows to change favourably when disturbed by external agents. The interest in the subject is steadily growing as the viscous action exerted by turbulence causes dramatic energy losses in flow systems of technological relevance, such as oil and gas pipelines, high-speed aircraft wings, jet engines intakes, and turbine blades. Even a small reduction of turbulence activity, and thus of wall friction, translates into improved system efficiency and therefore into lower fuel consumption. Further potential advantages are the attenuation of noise, structural vibrations, and aerodynamic heating.

Active turbulent drag reduction techniques, for which energy is introduced into the system, have received widespread attention, owing to the (so far) limited performances achieved by most passive techniques. Closed-loop feedback control strategies represent an emerging field of research where the activation is usually applied at the wall as wall-normal distributed transpiration \citep{kim-bewley-2007,kasagi-suzuki-fukagata-2009}. Open-loop techniques, for which the control law is predetermined, usually operate at much larger spatio-temporal scales and do not require distributed sensing. As for this type of forcing, near-wall flows have been excited by Lorentz forces or Dielectric Barrier Discharge plasma actuators, alternating wall suction and blowing, unsteady cross-flow pressure gradients and different types of wall motion with the intent to disrupt the self-sustaining turbulence production mechanisms. With regard to the space-time distribution of forcing, both spanwise- and streamwise-traveling waves have been employed. A recent volume \citep{leschziner-choi-choi-2011} contains several contributions to the subject.

We consider here the simplest amongst such open-loop techniques, i.e. the harmonic spanwise wall oscillations introduced by \cite{jung-mangiavacchi-akhavan-1992}. The oscillating wall has been chosen because it can be regarded as a paradigm for a larger class of drag-reduction techniques and because it offers the largest amount of available experimental and numerical data and the smallest number of forcing parameters. This flow has been studied mainly through turbulence statistics, flow visualizations of the near-wall modified flow, and simplified models which attempt to explain the physics behind drag reduction. Various mechanism for drag reduction have been proposed, such as the relative displacement of near-wall structures \citep{baron-quadrio-1996} and the creation of negative spanwise vorticity during the oscillation cycle \citep{choi-debisschop-clayton-1998}. In spite of such efforts, the answer to fundamental questions, such as of why the turbulent kinetic energy and the friction drag decrease and how the wall forcing can be modified most efficiently to achieve the largest net energy saving, still remain elusive.

The objective of the present work is therefore to gain further insight into the physics of an incompressible channel flow with spanwise wall oscillations. Although our conclusions will be limited to the oscillating wall, it is our hope that they will bear some generality. The focus is on how the energy transfer between the mean flow and the turbulent fluctuations is affected by the wall motion and on the role played by the forcing on the modification of the turbulent enstrophy. The approach is to identify those terms in the equations which are directly affected by the spanwise forcing and to single out the dominant one(s). Another important point is to study the energy transfer during the temporal evolution from the start-up of the wall motion with the aim of explaining the decrease of skin-friction coefficient. As statistical and flow visualization studies on drag-reducing flows are affected by the reference quantities used for dealing with dimensionless quantities, an important and critical choice made in this study is to carry out the numerical simulations with a constant streamwise pressure gradient: this provides us with a clear and unequivocal inner-units scaling. 

The flow configuration, the numerical procedures, the flow field decompositions and the basic flow statistics are presented in \S\ref{sec:numerical}. The analysis of the energy budget is given in \S\ref{sec:energy}. The turbulent enstrophy budget, chosen as a convenient substitute for the turbulent dissipation budget, is discussed in \S\ref{sec:enstrophy}. Section \S\ref{sec:summary} contains a summary of the results.

%------------------------------------
\section{Flow configuration and numerical procedures}
\label{sec:numerical}

An incompressible fully-developed turbulent channel flow between two indefinite parallel flat plates at a distance $2h^*$, driven by a constant streamwise pressure gradient $\Pi^*$, is studied by direct numerical simulations (DNS). Dimensional quantities are indicated by the symbol *. The coordinates $x^*$, $y^*$, $z^*$ indicate the streamwise, wall-normal and spanwise direction, respectively. The two walls at $y^*=0$ and $y^*=2h^*$ oscillate in phase along $z^*$ according to $w^*(x^*,z^*,t^*) = A^* \cos \left( 2 \pi t^*/T^* \right),$ where $t^*$ is time and $T^*$ is the oscillation period. Quantities are scaled by viscous units, i.e. by the kinematic viscosity of the fluid $\nu^*$ and the friction velocity $u_\tau^* = \sqrt{\tau_w^*/\rho^*}$, where $\tau_w^*$ is the time- and space-averaged wall-shear stress and $\rho^*$ is the density of the fluid. The friction velocity Reynolds number is $Re_\tau = u_\tau^* h^*/\nu^*=200$. As $\Pi^*$ is constant, the momentum balance at the walls shows that, once the oscillating-wall regime is established, $\tau_w^*$ (and therefore $Re_\tau$) retains the fixed-wall value. It follows that a unique wall-unit scaling is defined. Since this is the only scaling used throughout the paper, we omit the customary symbol $+$ marking inner-scaled quantities.  (We would like to point out here that the advantage in carrying out DNS at constant pressure gradient is not general. There is a potential impact on the computing costs for cases, like the present one, where an abrupt change of one parameter is introduced. As a general rule, a constant flow rate allows the wall friction to reach the new state sooner, thus yielding a shorter transient. However, this advantage is compensated by the integration time required to obtain a reliable value of the mean wall friction, which is averaged over two spatial directions only at each time step. This quantity thus presents larger temporal fluctuations than the flow rate, which is a volume-averaged quantity. At the present value of the Reynolds number, the computational cost of the two approaches is comparable.)

Details on the DNS code are found in \cite{luchini-quadrio-2006}. The computational domain has dimensions of $L_x^*=6 \pi h^*$, $L_y^*=2 h^*$, $L_z^*=3 \pi h^*$ in the three directions. The wall-normal direction is discretized by 160 mesh points and $320 \times 320$ Fourier modes are used along the homogeneous $x^*$ and $z^*$ directions. The time step is $\Delta t=0.1$ which guarantees that the CFL condition is amply verified for the chosen time integration scheme (a three-substep low-storage Runge-Kutta). The mean velocity profile and the variance of velocity fluctuations for the fixed-wall case, shown in figures~\ref{fig:DR} and \ref{fig:rms}, have been compared with those by \cite{delalamo-jimenez-2003} at a slightly lower Reynolds number and excellent agreement has been found.
When the wall oscillates, the amplitude (maximum speed) of the wall motion is $A=12$. The effect of $A$ on drag reduction has been previously studied by \cite{quadrio-ricco-2004} and is not considered here. The calculations span the range $0 \leq T < 175$, and most of the paper discusses one case with $T=100$, which is the oscillation period unless otherwise indicated.  

%------------------------------------
\subsection{Averaging operators and flow field decomposition}

This paper employs different types of space and time averages and the relevant operators are presented here. A quantity $f(x,y,z,t)$ is averaged along the homogeneous $x$ and $z$ directions as
\begin{equation*}
\averxz{f}(y,t)=\frac{1}{L_x L_z} \displaystyle \int_0^{L_x}\int_0^{L_z} f (x,y,z,t)\dif z \dif x.
\end{equation*}
The velocity and the vorticity fields, ${\bf U}={\bf U}(x,y,z,t)$ and ${\bf \Omega}={\bf \Omega}(x,y,z,t)$, are decomposed as follows
\[
{\bf U} = \left\{\averxz{U}(y,t),0,\averxz{W}(y,t)\right\} + \{u,v,w\}, \ \ {\bf \Omega} = \left\{\averxz{\Omega}_x(y,t),0,\averxz{\Omega}_z(y,t)\right\} + \{\omega_x,\omega_y,\omega_z\},
\]
where $\averxz{\Omega}_x= (1/2)\p \averxz{W}/\p y $ and $\averxz{\Omega}_z = -(1/2)\p \averxz{U}/\p y$ since $\averxz{V}=0$. At statistically steady state, a quantity $\averxz{f}(y,t)$ is averaged over the $N$ periods of oscillation as follows
\begin{equation*}
\averper{f}(y,\tau)=\frac{1}{N} \displaystyle\sum_{n=0}^{N-1} \averxz{f}(y,nT + \tau).
\end{equation*}
Note that these averaged quantities henceforth depend on the `window' phase-average time $\tau$, $0 \leq \tau < T$. Alternatively, they can be observed as a function of the oscillation phase $\phi=2 \pi \tau / T$. A quantity $f(y,\tau)$ is then averaged over $T$ according to
\[
\averph{f} (y) =\frac{1}{T} \int_0^T f (y,\tau) \dif \tau.
\]
A {\em global} quantity $\averglob{f}$ is obtained by integrating $\averph{f}(y)$ along $y$, as follows
\[
\averglob{f} = \int_0^h \averph{f} (y) \dif y.
\]
A transport equation is defined as global when its terms are global. All statistical samples are doubled by averaging over the two channel halves, by properly accounting for the existing symmetries.

%--------------------------------------------------
\subsection{Definition of turbulent drag reduction}
\label{sec:$DR$}

The skin-friction coefficient is defined as $C_f = 2 \tau_w^*/\rho^* U_b^{* 2}$, where $U_b^*$ is the bulk velocity,
\begin{equation}
\label{eq:ub}
U_b^*=\frac{\averglob{\averper{U}^*}} {h^*}.
\end{equation}
Following \cite{kasagi-hasegawa-fukagata-2009}, the drag reduction $R$ is defined as the change of $C_f$ with respect to the fixed-wall value $C_{f,0}$, i.e. $R = (C_{f,0} - C_f)/C_{f,0}$. When $\Pi^*$ is constant, $R$ is due to the increase of mass flow rate. As 
\begin{equation}
\label{eq:cf-ub}
C_f = \frac{2}{U_b^2},
\end{equation} 
$R$ may be written as
\begin{equation}
\label{eq:DR-def}
R = \frac{U_b^2 - U_{b,0}^2}{U_b^2}.
\end{equation}

%-----------------------------------
\subsection{Basic flow statistics}
\label{sec:ts}

\begin{figure}
  \centering
  \psfrag{x}{$y$}
  \psfrag{y}{$\averph{\averper{U}}$}
  \psfrag{D}{$R$}
  \psfrag{T}{$T$}
  \includegraphics[width=\columnwidth]{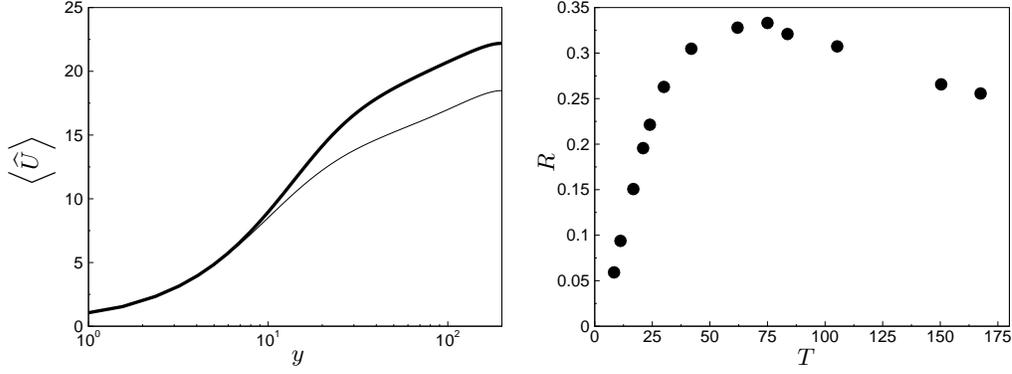}
  \caption{Left: Wall-normal profiles of $\averph{\averper{U}}$ for fixed-wall (thin line) and oscillating-wall (thick line) conditions. Right: $R$ as a function of $T$.}
  \label{fig:DR}
\end{figure}

Figure \ref{fig:DR} (left) shows that the mean velocity profile $\averph{\averper{U}}$ increases significantly throughout the channel for $T=100$ ($R=0.31$), while the wall-shear stress remains constant, in agreement with experimental studies where the drag-reduced friction velocity was used for inner scaling \citep{choi-debisschop-clayton-1998,ricco-wu-2004}. Figure \ref{fig:DR} (right) shows that $R$ increases sharply with $T$ up to the optimum $T_{opt} \approx 70$ and then decays at a slower rate. This behaviour is well documented by previous numerical studies, although quantitative differences can be ascribed to different scaling procedures. For example, the optimum period at constant $A^*$ is typically reported to be $T_{opt} \approx 100-125$ at constant mass flow rate if the fixed-wall $u_\tau^*$ is used for scaling.

\begin{figure}
  \centering
  \psfrag{x}{$y$}
  \psfrag{y}{$\averph{\averper{u^2}}$,$\averph{\averper{v^2}}$,$\averph{\averper{w^2}}$,$\averph{\averper{uv}}$}
  \psfrag{k}{$y$}
  \psfrag{l}{$\averper{vw}$}
  \psfrag{a}{\tiny $\phi=0$}
  \psfrag{b}{\tiny $\phi=\frac{\pi}{4}$}
  \psfrag{c}{\tiny $\phi=\frac{\pi}{2}$}
  \psfrag{d}{\tiny $\phi=\frac{3\pi}{4}$}
  \includegraphics[width=\columnwidth]{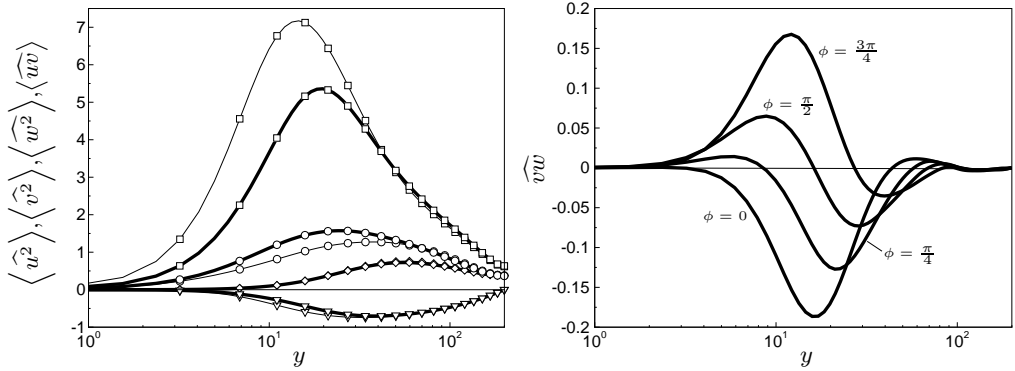}
  \caption{Left: Wall-normal profiles of the variance of velocity fluctuations and of Reynolds stress $\averph{\averper{uv}}$ for fixed-wall (thin lines) and oscillating-wall (thick lines) cases. Squares: $u$, diamonds: $v$, circles: $w$, triangles: $\averph{\averper{uv}}$. Right: Wall-normal profiles of Reynolds stresses $\averper{v w}$ at different values of the phase angle $\phi$.}
 \label{fig:rms}
\end{figure}

The variance of the turbulent velocity fluctuations and the Reynolds stress component $\averph{\averper{uv}}$ are shown in figure \ref{fig:rms} (left). The wall motion primarily affects $\averph{\averper{u^2}}$ up to $y \approx 30$; the peak decreases and its position shifts upward from $y \approx 14$ to $y \approx 20$. The profile of $\averph{\averper{v^2}}$ is largely unvaried, while that of $\averph{\averper{w^2}}$ increases up to $y \approx 40$. As discussed by \cite{quadrio-ricco-2011} in the context of streamwise-travelling waves, it appears that the large reductions of turbulence fluctuations for all the velocity components often reported in the literature are largely a byproduct of the outer scaling employed to compare flows that in fact have different values of $Re_\tau$ owing to drag reduction. The Reynolds stresses $\averph{\averper{uv}}$ are attenuated up to $y \approx 60$. This is consistent with \cite{marusic-joseph-mahesh-2007}'s finding on the relation between drag reduction and a weighted integral of $\averph{\averper{uv}}$, an extension of the result by \cite{fukagata-iwamoto-kasagi-2002} to the $\Pi$-constant case.

The wall oscillation induces the additional phase-varying Reynolds stresses $\averper{vw}$, shown in figure \ref{fig:rms} (right). This term is null in the fixed-wall case. At opposite phases of the cycle, the $\averper{vw}$ profiles show the same behaviour with opposite sign, which leads to $\averph{\averper{vw}}=0$.

\section{Energy balance}
\label{sec:energy}

As follows from (\ref{eq:cf-ub}), the reduction of $C_f$ can be understood by studying how $U_b$ increases. As a first step, we therefore study the transport equations for the mean kinetic energy (MKE),$\left(\averper{U}^2+\averper{W}^2\right)/2$, where $U_b$ appears explicitly, and for the turbulent kinetic energy (TKE), $\averper{q^2}/2$, where $q^2=u_i u_i$. (The Einstein summation convention of repeated indices is adopted henceforth and the subscripts $i=1,2,3$ denote the $x,y,z$ directions and the corresponding velocity and vorticity components.) These two equations are then summed to obtain the global balance for the total kinetic energy. 

%---------------------------
\subsection{Mean kinetic energy balance}
The transport equation for MKE reads
\begin{equation}
\label{eq:mke-inst}
\begin{split}
  \underbrace{\frac{1}{2} \frac{\p \left(\averper{U}^2 + \averper{W}^2\right)}{\p \tau}}_1 +
  \underbrace{\averper{U}\Pi}_2 &=
  \underbrace{\averper{uv}\frac{\p \averper{U}}{\p y}}_3 +
  \underbrace{\averper{v w}\frac{\p \averper{W}}{\p y}}_4 -
  \underbrace{\frac{\p \left(\averper{uv}\averper{U}\right)}{\p y}}_5 -
  \underbrace{\frac{\p \left(\averper{v w}\averper{W}\right)}{\p y}}_6 \\ &+
  \underbrace{\frac{\p}{\p y}\left(\averper{U} \ \frac{\p \averper{U}}{\p y}\right)}_7 +
  \underbrace{\frac{\p}{\p y}\left(\averper{W} \ \frac{\p \averper{W}}{\p y}\right)}_8 -
  \underbrace{\left(\frac{\p \averper{U}}{\p y}\right)^2}_9 -
  \underbrace{\left(\frac{\p \averper{W}}{\p y}\right)^2}_{10}.
\end{split}
\end{equation}
Term 1 denotes the temporal change of MKE and term 2 is the work per unit time done by $\Pi$, i.e. the power used to drive the flow along the $x$ direction. Thanks to the wall oscillation, the system absorbs more kinetic energy than in the fixed-wall case through the increment of $\averper{U}$. Term 3 is the work of deformation carried out by the Reynolds stresses $\averper{uv}$, through which energy is exchanged between the mean flow and the fluctuating flow. Term 4 indicates the work of deformation done by the Reynolds stresses $\averper{v w}$; similarly to term 3, it transfers energy between the mean flow and the fluctuating flow. Terms 3 and 4 appear with opposite sign in the TKE equation, as shown in \S\ref{sec:tke}. The transport works performed by the Reynolds stresses $\averper{uv}$ and $\averper{v w}$ are described by terms 5 and 6, respectively. Terms 7 and 8 are the transport works done by the mean streamwise and spanwise viscous stresses, respectively. Term 9 is the viscous dissipation of MKE by the wall-normal gradient of $\averper{U}$, while term 10 is the viscous dissipation by the wall-normal gradient of $\averper{W}$.

\begin{figure}
  \centering
  \psfrag{x}{$y$}
  \psfrag{y}{$-\averper{v w} \p \averper{W} / \p y$}
  \psfrag{a}{\scriptsize $\phi=0$}
  \psfrag{b}{\scriptsize $\phi=\frac{\pi}{4}$}
  \psfrag{c}{\scriptsize $\phi=\frac{\pi}{2}$}
  \psfrag{d}{\scriptsize $\phi=\frac{3\pi}{4}$}
  \includegraphics[width=\columnwidth]{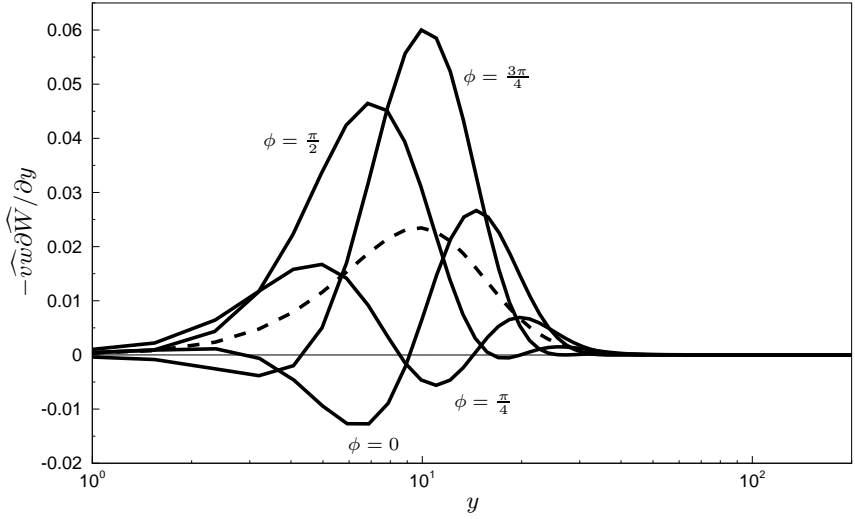}
  \caption{Wall-normal profiles of $-\averper{v w} \p \averper{W} / \p y$ at different phases $\phi$ (solid lines) and their time-averaged value (dashed line).}
\label{fig:Pvw}
\end{figure}
The second part of term 1 and terms 4, 6, 8, 10 are directly related to the wall oscillation, since $\averper{W}$ appears explicitly in their expressions. The turbulent production term 4, $-\averper{v w} \p \averper{W} / \p y$, which is absent in the fixed-wall case because $\averper{v w}$ and $\averper{W}$ are null, is shown in figure \ref{fig:Pvw} at different phases of the cycle. Although it is negative during part of the cycle (mainly for $y<15$, when it instantaneously extracts energy from the turbulent fluctuations to enhance MKE), it is positive for most of the cycle, i.e. its average contribution is to transfer MKE to the turbulent fluctuations; see dashed line in figure \ref{fig:Pvw}, which represents $- \averph{ \averper{v w} \p \averper{W} / \p y }$.

As the primary interest resides in the change of $U_b$, the MKE equation (\ref{eq:mke-inst}) is now time-averaged and integrated along $y$ to make $U_b$ appear in the energy balance. Time averaging eliminates term 1 because of time periodicity. Terms 2, 3, 4, 9, 10 are retained as is term 8 since $\averper{W}$ is non-zero at $y=0$. Terms 5 and 6 disappear because $\averper{uv}$ and $\averper{v w}$ are null at $y=0$ and at $y=h$. Term 7 becomes null because $\averper{U}=0$ at $y=0$ and $\p \averper{U} / \p y=0$ at $y=h$.

The global transport equation for MKE is 
\begin{equation}
\begin{split}
\label{eq:mke-integral}
  U_b \tau_w +
  \underbrace{\averph{A \pipe{\frac{\p \averper{W}}{\p y}}_{y=0}}}_{\mathcal{E}_w} =  -
  \underbrace{\averglob{\averper{uv}\frac{\p \averper{U}}{\p y}}}_{\mathcal{P}_{uv}} -
  \underbrace{\averglob{\averper{v w}\frac{\p \averper{W}}{\p y}}}_{\mathcal{P}_{vw}}  +
  \underbrace{\averglob{\left(\frac{\p \averper{U}}{\p y}\right)^2}}_{\mathcal{D}_{U}} +
  \underbrace{\averglob{\left(\frac{\p \averper{W}}{\p y}\right)^2}}_{\mathcal{D}_{W}},
\end{split}
\end{equation}
where $\tau_w = \averph{\p \averper{U}/\p y|_{y=0}}$.
The first term on l.h.s. comes from term 2 in (\ref{eq:mke-inst}) and represents the global energy per unit time pumped into the system through the external pressure gradient $\Pi$. Term $\mathcal{E}_w$ is the energy input given by the wall motion, and denotes the energy spent to move the walls against the frictional resistance of the fluid. It stems from the transport term 8 in (\ref{eq:mke-inst}). Terms $\mathcal{P}_{uv}$ and $\mathcal{P}_{vw}$, which originate from terms 3 and 4 in (\ref{eq:mke-inst}), are a sink for MKE and appear in the global TKE balance as production terms. Terms $\mathcal{D}_{U}$ and $\mathcal{D}_{W}$, which stem from terms 9 and 10 in (\ref{eq:mke-inst}) respectively, denote the global viscous dissipation due to the gradients of the mean streamwise and spanwise velocity components. Equation (\ref{eq:mke-integral}) represents the first step toward understanding drag reduction because $U_b$ now appears explicitly. It states that part of the energy input, $U_b \tau_w + \mathcal{E}_w$, is transferred to the turbulence via $\mathcal{P}_{uv}$ and $\mathcal{P}_{vw}$, and the remaining part is dissipated into heat through $\mathcal{D}_{U}$ and $\mathcal{D}_{W}$.

%------------------------------------
\subsection{Turbulent kinetic energy balance}
\label{sec:tke}

The  transport equation for TKE reads
\begin{equation}
  \label{eq:tke-inst}
  \underbrace{\frac{1}{2}\frac{\p \averper{q^2}}{\p \tau}}_1 = 
  \underbrace{-\frac{\p \left(\averper{v p}\right)}{\p y} - \frac{1}{2} \frac{\p \left(\averper{v q^2}\right)}{\p y}}_2 -
  \underbrace{\averper{uv}\frac{\p \averper{U}}{\p y}}_3 -
  \underbrace{\averper{v w}\frac{\p \averper{W}}{\p y}}_4 +
  \underbrace{\frac{1}{2} \frac{\p^2 \averper{q^2}}{\p y^2}}_5 -
  \underbrace{\averper{\frac{\p u_i}{\p x_j} \frac{\p u_i}{\p x_j}}}_6,
\end{equation}
where $p$ is the turbulent pressure. The temporal change of TKE is expressed by term 1, while terms 2 represent the work of transport done by the total dynamic pressure of turbulence. Terms 3 and 4 denote production of TKE and also appear in the MKE equation (\ref{eq:mke-inst}) with opposite sign. Terms 5 and 6 together represent the combined effect of the work done by the viscous shear stresses of the turbulent motion and of the viscous dissipation of TKE into heat. Term 6 is often referred to as the pseudo-dissipation (see \cite{pope-2000} at page 132). The turbulent production term 4, $\averper{v w} \p \averper{W} / \p y$, is the only one containing $\averper{W}$ explicitly.

Analogously to the analysis of the MKE equation, time averaging and integration along $y$ lead to the following simplifications. Term 1 disappears because of time periodicity. Terms 2 become null upon $y$-integration because of the no-slip condition at $y=0$ and $\averper{v p}=\averper{v q^2}/2=0$ at $y=h$. Term 5 is also null because
\[
 \averglob{\frac{\p^2 \averper{q^2}}{\p y^2}} =
 \int_0^h \frac{\p}{\p y}\averph{\frac{\p \averper{q^2}}{\p y}} \dif y =
 \pipe{\averph{\frac{\p \averper{q^2}}{\p y}}}_{y=h} -
 \pipe{2\averph{\averper{q \frac{\p q}{\p y}}}}_{y=0}=0,
\]
as $\p \averper{q^2} / \p y=0$ at $y=h$ and $q=0$ at $y=0$.

The global transport equation for TKE is
\begin{equation}
\label{eq:tke-integral}
  \underbrace{\averglob{\averper{uv}\frac{\p \averper{U}}{\p y}}}_{\mathcal{P}_{uv}}
  +\underbrace{\averglob{\averper{v w}\frac{\p \averper{W}}{\p y}}}_{\mathcal{P}_{vw}}
  +\averglob{\averper{\frac{\p u_i}{\p x_j} \frac{\p u_i}{\p x_j}}} = 0,
\end{equation}
where $\mathcal{P}_{uv}$ and $\mathcal{P}_{vw}$ are as in (\ref{eq:mke-integral}). The next-to-last equation at page 74 in \cite{hinze-1975} shows that the last term in (\ref{eq:tke-integral}) is the global TKE dissipation,
\begin{equation}
\label{eq:tke-diss}
 \mathcal{D}_\turb \equiv
 \averglob{\averper{\frac{\p u_i}{\p x_j}\left(\frac{\p u_i}{\p x_j} + \frac{\p u_j}{\p x_i}\right)}} =
 \averglob{\averper{\frac{\p u_i}{\p x_j} \frac{\p u_i}{\p x_j}}}.
\end{equation}
Equation (\ref{eq:tke-integral}) may therefore be written as
\begin{equation}
\label{eq:tke-global}
-\mathcal{P}_{uv} - \mathcal{P}_{vw} = \mathcal{D}_\turb.
\end{equation}

\begin{figure}
  \centering
  \psfrag{x}{$y$}
  \psfrag{y}{}
  \includegraphics[width=\columnwidth]{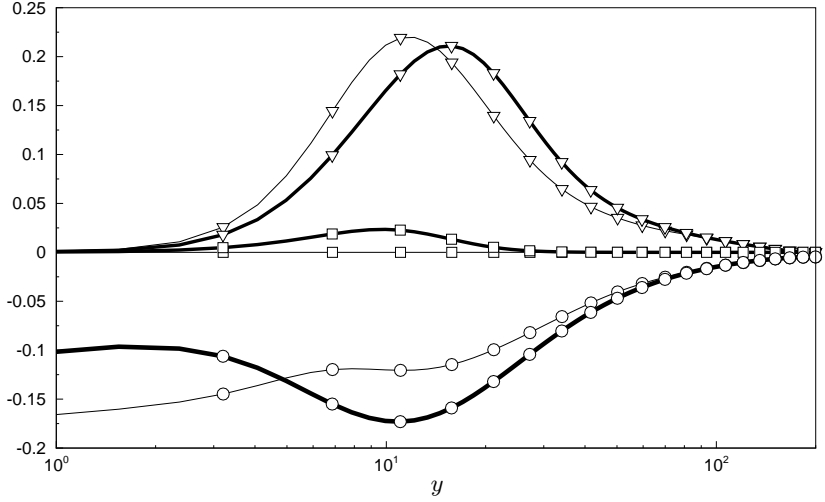}
  \caption{Wall-normal profiles of the integrands of $\mathcal{P}_{uv}$, $\mathcal{P}_{vw}$, and $\mathcal{D}_\turb$ in (\ref{eq:tke-global}). Time-averaged values of production term 3 in (\ref{eq:tke-inst}), denoted by triangles, of production term 4 in (\ref{eq:tke-inst}), denoted by squares, and of pseudo-dissipation term 6 in (\ref{eq:tke-inst}), denoted by circles, for fixed-wall (thin lines) and oscillating-wall (thick lines) cases.}
  \label{fig:budget_tke}
\end{figure}

The balance in (\ref{eq:tke-global}) simply states that the global TKE engendered by the production terms, $\mathcal{P}_{uv}$ and $\mathcal{P}_{vw}$, is dissipated into heat by the turbulent viscous stresses. Figure \ref{fig:budget_tke} shows the wall-normal profiles of the three terms whose integrals compose the balance (\ref{eq:tke-global}). It is observed that the integrand of $\mathcal{P}_{uv}$ is suppressed near the oscillating wall and its peak moves upward. These changes are attributed to the increase of $\dif \averph{\averper{U}}/ \dif y$ in the outer region and to the near-wall reduction of $\averph{\averper{uv}}$,  as shown in figures \ref{fig:DR} (left) and \ref{fig:rms} (left), respectively. In contrast, $\mathcal{D}_\turb$ decreases near the wall, but it significantly increases at $y \approx 10$. As shown in \S\ref{sec:enstrophy}, this is directly linked to the enstrophy production through the stretching of vorticity fluctuations by the Stokes layer. 

%--------------------------------------------
\subsection{Total kinetic energy balance}
\label{sec:total}
By summing the global transport equations for MKE, (\ref{eq:mke-integral}), and TKE, (\ref{eq:tke-integral}), the global balance for the total mechanical energy is found
\begin{equation}
\label{eq:total}
U_b \tau_w + \mathcal{E}_w = \mathcal{D}_U + \mathcal{D}_W + \mathcal{D}_\turb.
\end{equation}
The energy input $U_b \tau_w$ (per unit area and unit time), which drives the flow along $x$, and the energy $\mathcal{E}_w$, spent to enforce the wall motion, are dissipated into heat through the viscous action of the mean streamwise and spanwise flow gradients, denoted by $\mathcal{D}_U$ and $\mathcal{D}_W$ respectively, and through the viscous dissipation $\mathcal{D}_\turb$ of the turbulent fluctuations. Note that, as shown by \cite{laadhari-2007} for the uncontrolled flow, $\mathcal{D}_\turb \gg \mathcal{D}_U$ as $Re_\tau \rightarrow \infty$.

\begin{figure}
  \centering
  \psfrag{DU}{$\mathcal{D}_U$}
  \psfrag{DW}{$\mathcal{D}_W$}
  \psfrag{DT}{$\mathcal{D}_\turb$}
  \psfrag{Puv}{$\mathcal{P}_{uv}$}
  \psfrag{Pvw}{$\mathcal{P}_{vw}$}
  \psfrag{Ub}{$U_b \tau_w$}
  \psfrag{Sw}{$\mathcal{E}_w$}
  \psfrag{Mx}{MKE-$x$}
  \psfrag{Mz}{MKE-$z$}
  \psfrag{T}{TKE}
  \psfrag{Ub-ow}{\plus 3.5}
  \psfrag{Ub-ref}{15.9}
  \psfrag{Sw-ow}{\plus 13.2}
  \psfrag{Dw-ow}{\plus 12.9}
  \psfrag{Du-ow}{\plus 2.7}
  \psfrag{Du-ref}{9.4}
  \psfrag{Puv-ow}{\plus 0.8}
  \psfrag{Puv-ref}{6.5}
  \psfrag{Pvw-ow}{\plus 0.3}
  \psfrag{Dt-ow}{\plus 1.1}
  \psfrag{Dt-ref}{6.5}
  \includegraphics[width=\columnwidth]{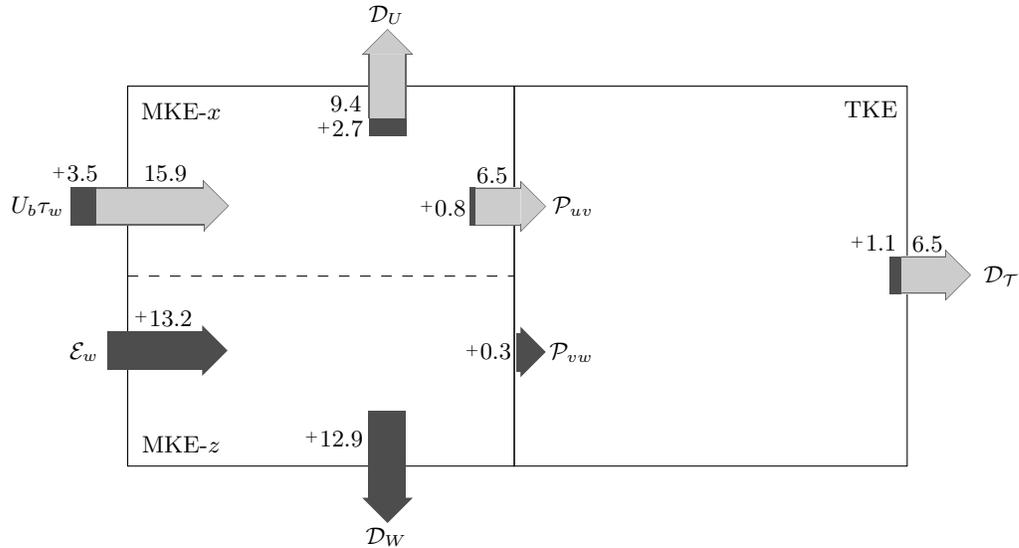}
  \caption{Schematic of the global energy balance for the total mechanical energy. The numbers indicate the magnitude of the terms and the additional contribution due to the wall oscillation. The light grey portions of arrows denote the contributions at fixed-wall conditions, while the dark grey arrows or portions of arrows indicate the changes due to the wall motion.}
  \label{fig:energy-balance}
\end{figure}

Figure \ref{fig:energy-balance} summarizes and quantifies the global energy balance. The two boxes represent MKE and TKE; MKE-$x$ and MKE-$z$ indicate the portion of the MKE balance pertaining to the streamwise and spanwise directions, respectively. The light grey portions of arrows indicate the energy terms in the fixed-wall case, while the dark grey arrows or portions of arrows denote the energy transfers due to the wall motion. The schematic graphically highlights that the production terms $\mathcal{P}_{uv}$ and $\mathcal{P}_{vw}$ only transfer energy ``internally'' between MKE and TKE, therefore disappearing from the total energy balance (\ref{eq:total}).

As $C_f$ and $U_b$ are related, the aim is to study how the wall motion acts on $U_b$ to discern information on drag reduction. The total energy balance (\ref{eq:total}) is therefore analyzed in more detail because it contains $U_b$ explicitly. As shown in figure \ref{fig:energy-balance}, it is first noted that the terms in (\ref{eq:mke-integral}) pertaining to the streamwise and spanwise directions are decoupled, so that 
\begin{equation}
\label{eq:MKE-x}
U_b \tau_w = \mathcal{P}_{uv} + \mathcal{D}_U,
\end{equation}
and, correspondingly, $\mathcal{E}_w = \mathcal{P}_{vw} + \mathcal{D}_W$. The two terms containing $\averper{W}$, i.e. $\mathcal{E}_w$ and $\mathcal{D}_W$, almost balance each other; the difference, given by $\mathcal{P}_{vw}$ (which is absent in (\ref{eq:total})), is much smaller than the other terms in (\ref{eq:total}). To gain insight into the changes of $U_b$, one is thus led to investigate how the wall oscillation affects the dynamics of the two remaining relevant terms, i.e. $\mathcal{D}_U$ and $\mathcal{D}_\turb$. 
The relative contribution of $\mathcal{D}_U$ to the input power $U_b \tau_w$ increases in the oscillating-wall case. In the fixed-wall case, the input power in viscous units is 15.9, 59\% (i.e. 9.4/15.9) of which is dissipated by $\mathcal{D}_U$. When the wall oscillates, this share increases to 62\% (i.e. 12.1/19.4). This fact agrees with previous studies on flow control \citep{bewley-2009,fukagata-sugiyama-kasagi-2009}, which show that, as $C_f$ decreases as the flow tends to the laminar regime, the input power is dissipated more by $\mathcal{D}_U$ and less by $\mathcal{D}_\turb$. 

Three different scenarios might explain how the wall motion acts on $\mathcal{D}_U$ and why the relative contribution of $\mathcal{D}_U$ in the global balance increases during the wall motion. 

\begin{enumerate}
\item In the first scenario, the mean spanwise shear may work directly on $\mathcal{D}_U$. The transport equation for $\left(\p\averper{U}/\p y\right)^2$, the integrand of $\mathcal{D}_U$ (see (\ref{eq:mke-integral})), is thus studied. It reads
\begin{equation}
\label{eq:dUdy-2}
  \frac{1}{2}\frac{\p}{\p \tau}\left[\left(\frac{\p\averper{U}}{\p y}\right)^2\right] = -
  \frac{\p^2\averper{uv}}{\p y^2}\frac{\p\averper{U}}{\p y} +
  \frac{1}{2}\frac{\p^2}{\p y^2}\left[\left(\frac{\p\averper{U}}{\p y}\right)^2\right] -
  \frac{\p^2\averper{U}}{\p y^2} .
\end{equation}
The spanwise velocity $\averper{W}$ does not appear in the transport equation for $(\p\averper{U}/\p y)^2$, which proves that the oscillating wall does not influence the dynamics of the mean streamwise flow directly and that the increase of $\mathcal{D}_U$ is linked to the modification of turbulent dynamics. 

\item In the second scenario, the spanwise viscous effects directly damp $\mathcal{P}_{uv}$, so that the relative contribution of $\mathcal{D}_U$ in (\ref{eq:MKE-x}) is larger. As $\mathcal{P}_{uv}$ depends on $\averper{uv}$ and $\p\averper{U}/\p y$ (see (\ref{eq:mke-integral})) and the direct action of $\averper{W}$ on $\p\averper{U}/\p y$ has been excluded, the focus is on the transport equation for the Reynolds stresses $\averper{uv}$ (which also appear in the r.h.s. of (\ref{eq:dUdy-2}))
\begin{equation}
\label{eq:uv}
  \frac{\p \left(\averper{uv}\right)}{\p \tau} = - \averper{v^2} \frac{\p \averper{U}}{\p y} - \frac{\p \left(\averper{uv^2}\right)}{\p y} - \left( \averper{v \frac{\p p}{\p x}} +  \averper{u \frac{\p p}{\p y}} \right) + \frac{\p^2 \averper{uv}}{\p y^2} - \averper{ \frac{\p u}{\p x_j}\frac{\p v}{\p x_j} }.
\end{equation}
The mean flow $\averper{W}$ does not appear explicitly in (\ref{eq:uv}), which demonstrates that the oscillation does not work directly on $\averper{uv}$ either. Therefore, this scenario is excluded as $\mathcal{P}_{uv}$ is not immediately affected by the large-scale spanwise flow because neither $\averper{uv}$ nor $\p \averper{U}/{\p y}$ are.

\item The third scenario is rather counterintuitive. The wall oscillation enhances the turbulent dissipation, so that the turbulent activity drops due to the increased dissipative nature of the flow. A relatively lower level of dissipation of turbulent energy into heat thus sets in to balance this new condition. A similar behaviour has been observed in applying the suboptimal control theory \citep{lee-kim-choi-1998} when the turbulent dissipation is employed as a cost function. 

\end{enumerate}

Elucidating the third scenario, i.e. understanding how the wall motion affects the dynamics of the turbulent viscous dissipation $\mathcal{D}_\turb$, then becomes the aim of the next section.

\section{Turbulent enstrophy balance}
\label{sec:enstrophy}

By use of the fluctuating vorticity, term 6 in (\ref{eq:tke-inst}) becomes \citep{pope-2000}
\begin{equation}
\label{eq:diss-vort}
 \averper{\frac{\p u_i}{\p x_j} \frac{\p u_i}{\p x_j}} = 
 \averper{\omega_i \omega_i} + \frac{\p^2 (\averper{u_i u_j})}{\p x_i \p x_j}.
\end{equation}
The global dissipation of TKE in (\ref{eq:tke-inst}) becomes
\begin{equation}
\label{eq:dt-enstrophy}
\mathcal{D}_\turb = \averglob{\averper{\omega_i \omega_i}},
\end{equation}
which follows from the substitution of (\ref{eq:diss-vort}) into (\ref{eq:tke-diss}), from the periodicity along the homogeneous $x$ and $z$ directions, from the velocity fluctuations being zero at $y=0$ and because $\p\averper{v^2}/\p y=0$ at $y=h$. Note that the viscous dissipation of the total mechanical energy,  $\mathcal{D}_U$+$\mathcal{D}_W$+$\mathcal{D}_\turb$, equals the global enstrophy only in the case of stationary boundaries \citep{davidson-2004}, and therefore not in the oscillating-wall case. However, equation (\ref{eq:dt-enstrophy}) is valid for the wall-oscillation case because the turbulent fluctuations vanish at the walls.

Instead of considering the transport equation for the turbulent energy dissipation, we opt to study the turbulent enstrophy equation. In the second part of the Appendix, it is shown that the form of the turbulent dissipation equation is similar to that of the enstrophy equation and that the dominant terms brought about by the wall oscillation in these equations have the same order of magnitude. Expressing $\mathcal{D}_\turb$ in terms of the turbulent enstrophy is more compact than if the turbulent dissipation is used (compare (\ref{eq:dt-enstrophy}) with (\ref{eq:tke-diss})). The enstrophy equation has the further advantage over the dissipation equation that the turbulent pressure does not need to be computed. (This advantage is shared by the Orr-Sommerfeld and vorticity formulations of the Navier-Stokes equations over the framework involving primitive variables.) Moreover, the physical meaning conveyed by the enstrophy equation is arguably more immediate than the one provided by the dissipation equation; for example, terms 2 and 3 in the turbulent enstrophy equation (\ref{eq:enstrophy}) denote production of vorticity, while the corresponding terms in the dissipation equation indicate production of turbulent dissipation.

%--------------------------------------------------------
\subsection{Balance equation for the turbulent enstrophy}

The transport equation for the turbulent enstrophy \citep{tennekes-lumley-1972} reads
\begin{equation}
\label{eq:enstrophy}
 \begin{split}
 \underbrace{\frac{1}{2}\frac{\p\averper{\omega_i\omega_i}}{\p \tau}}_{1}=&
 \underbrace{\averper{\omega_x\omega_y}\frac{\p \averper{U}}{\p y}}_{2} +
 \underbrace{\averper{\omega_z\omega_y}\frac{\p \averper{W}}{\p y}}_{3} +
 \underbrace{\averper{\omega_j\frac{\p u}{\p x_j}}\frac{\p\averper{W}}{\p y}}_{4} -
 \underbrace{\averper{\omega_j\frac{\p w}{\p x_j}}\frac{\p\averper{U}}{\p y}}_{5} \\
 &-\underbrace{\averper{v\omega_x}\frac{\p^2\averper{W}}{\p {y}^2}}_{6} +
 \underbrace{\averper{v\omega_z}\frac{\p^2\averper{U}}{\p {y}^2}}_{7} +
 \underbrace{\averper{\omega_i\omega_j\frac{\p u_i}{\p x_j}}}_{8} -
 \underbrace{\frac{1}{2}\frac{\p}{\p y}\left(\averper{v\omega_i\omega_i}\right)}_{9} \\
 &+\underbrace{\frac{1}{2}\frac{\p^2\averper{\omega_i\omega_i}}{\p {y}^2}}_{10} -
 \underbrace{\averper{\frac{\p\omega_i}{\p x_j}\frac{\p\omega_i}{\p x_j}}}_{11}.
 \end{split}
\end{equation}
Term 1 indicates the time rate of change of the turbulent enstrophy. Terms 2 and 3 are the production (or removal) of turbulent vorticity caused by stretching (or squeezing) of vorticity fluctuations by the mean flow gradients $\p \averper{U} / \p y$ and $\p \averper{W} / \p y$, respectively. Terms 4 and 5 indicate the production of mean and turbulent enstrophy by the stretching of fluctuating vorticity through the fluctuating strain rates $\p u / \p x_j$ and $\p w / \p x_j$, respectively. Terms 6 and 7 represent the exchange of fluctuating vorticity between the mean and the turbulent enstrophy due to the gradients of streamwise and spanwise mean vorticity, respectively. They are analogous to the turbulent kinetic energy production terms in the MKE and TKE equations (\ref{eq:mke-inst}) and (\ref{eq:tke-inst}). Term 8 is the production of turbulent enstrophy by stretching of turbulent vorticity through turbulent velocity gradients. Term 9 denotes the transport of turbulent enstrophy by the fluctuating wall-normal velocity component. Term 10 is the viscous transport of turbulent enstrophy and term 11 is the viscous dissipation of turbulent enstrophy. The only terms in (\ref{eq:enstrophy}) that become null when \eqref{eq:enstrophy} is made global are term 1, when time averaged because of time periodicity, and term 9 when integrated along $y$.

In contrast to the case of the transport equation \eqref{eq:dUdy-2} for $\left(\p \averper{U} / \p y\right)^2$, which contributes to $\mathcal{D}_{U}$ (see \eqref{eq:mke-integral}), $\averper{W}$ appears explicitly in terms 3, 4 and 6 of \eqref{eq:enstrophy}. These terms arise only when the wall oscillates. This indicates that the spanwise motion acts directly on the turbulent enstrophy and therefore on the global turbulent dissipation $\mathcal{D}_\turb$. As this quantity increases during the wall motion, it is worth studying how these oscillating-wall terms contribute to modify the enstrophy balance and, in turn, $U_b$ through (\ref{eq:total}) and $C_f$ through (\ref{eq:cf-ub}).

\begin{figure}
  \centering
  \psfrag{x}{$y$}
  \psfrag{y}{}
  \psfrag{b}{\mymk{2}}
  \psfrag{c}{\mymk{3}}
  \psfrag{d}{\mymk{4}}
  \psfrag{e}{\mymk{5}}
  \psfrag{f}{\mymk{6}}
  \psfrag{g}{\mymk{7}}
  \psfrag{h}{\mymk{8}}
  \psfrag{i}{\mymk{9}}
  \psfrag{j}{\mymk{10}}
  \psfrag{m}{\mymk{11}}
  \includegraphics[width=\textwidth]{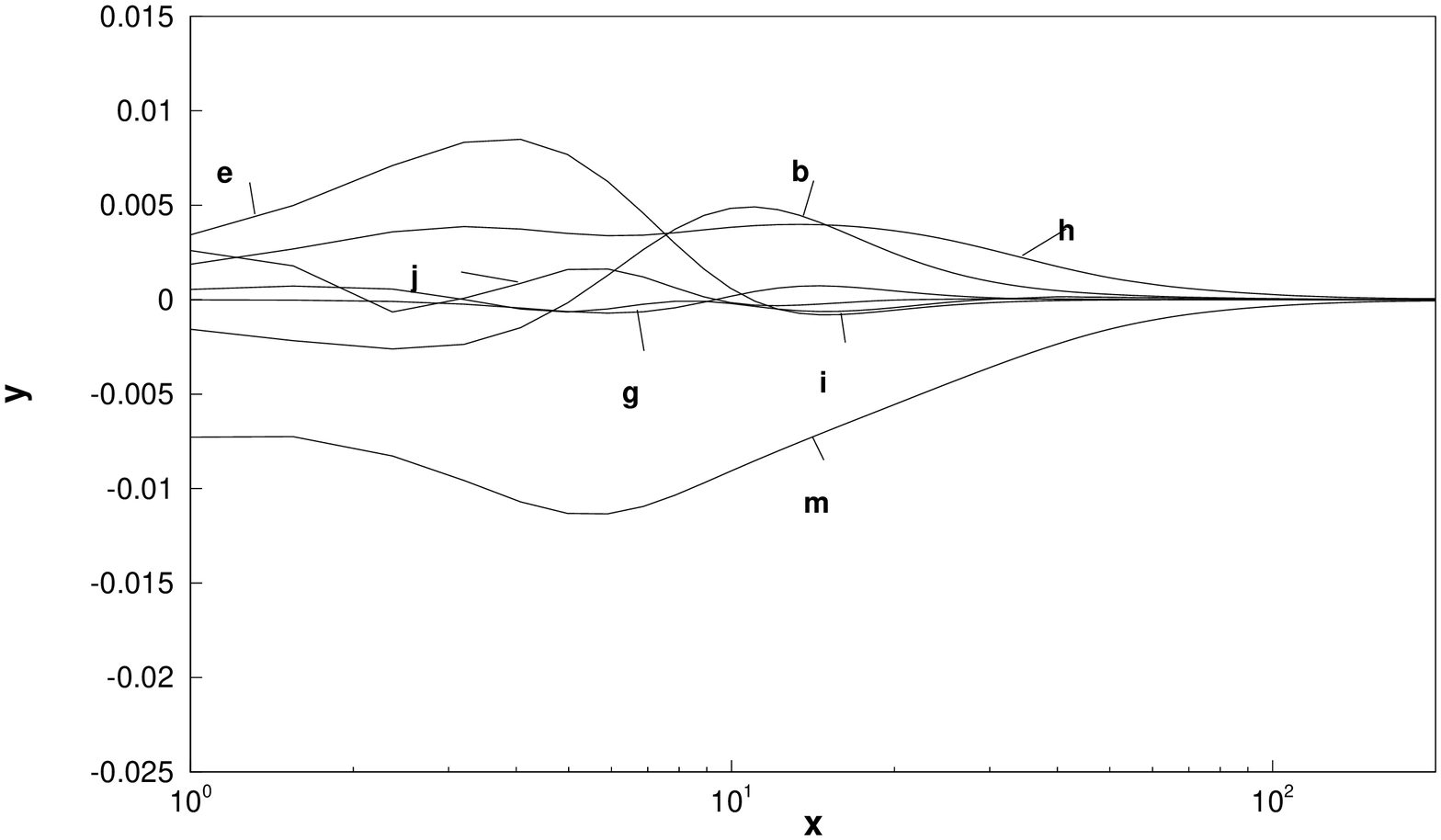} \\
  \includegraphics[width=\textwidth]{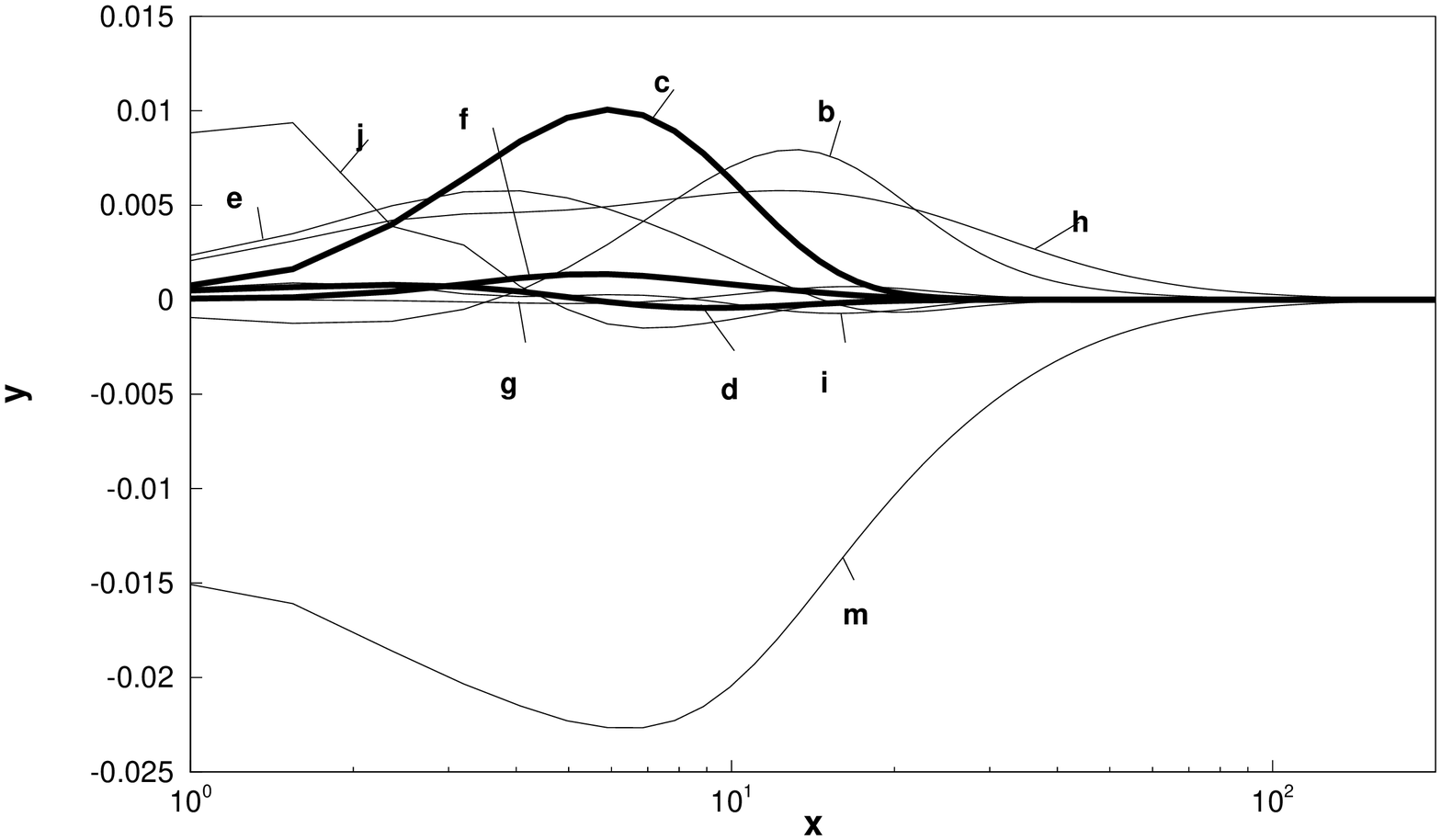}
  \caption{Wall-normal profiles of the time-averaged terms in the turbulent enstrophy equation (\ref{eq:enstrophy}) for the fixed-wall case (top) and the oscillating-wall case (bottom). Thick lines highlight terms only occurring during the wall motion.}
  \label{fig:budget-enstrophy}
\end{figure}

Figure \ref{fig:budget-enstrophy} shows the profiles of the time-averaged terms in the turbulent enstrophy balance (\ref{eq:enstrophy}) for the fixed-wall (top) and oscillating-wall (bottom) cases once the new fully-developed regime has established. The numbers refer to the terms in \eqref{eq:enstrophy} and the thick lines in the bottom graph highlight terms only occurring during the wall motion. The fixed-wall profiles show very good agreement with the ones in \cite{antonia-kim-1994}, \cite{gorski-wallace-bernard-1994} and \cite{abe-antonia-kawamura-2009}. (Note that in \cite{gorski-wallace-bernard-1994} and \cite{abe-antonia-kawamura-2009} the terms are multiplied by a factor of 2.) In the oscillating-wall case, the vorticity production term 3, $\averph{\averper{\omega_z\omega_y} \p \averper{W} / \p y}$, is dominant in the proximity of the wall, $y<10$, over terms 4 and 6, and over the production and transport terms already present in the fixed-wall case, i.e. terms 2, 5, 7, 8, 10. This is the key term producing turbulent enstrophy (and dissipation). Its physical meaning is further addressed in \S\ref{sec:physics}. It peaks at $y \approx 6$ and distinctly affects term 11, the dissipation of turbulent enstrophy, at the edge of the viscous sublayer and in the lower part of the buffer region, as clear from the similar shapes of the profiles for $2 < y < 20$.

In a very thin near-wall layer, $y < 2$, term 3 is small. Term 10, the viscous transport of turbulent enstrophy, is instead responsible for the intense increase of dissipation of turbulent enstrophy there. While the production term $\averph{\averper{\omega_z\omega_y} \p \averper{W} / \p y}$ emphasizes the direct action of the spanwise shear layer on the turbulent enstrophy, the increase of the production term 2, $\averph{\averper{\omega_x \omega_y} \p \averper{U} / \p y}$, outlines the indirect effect of the wall motion caused by the increment of $\p \averper{U} / \p y$.
Term 3 is primarily dominant near the wall, whereas term 2 increases at higher locations. This is because term 3 is dictated by the near-wall spanwise velocity $\averper{W}$, while $\p \averper{U} / \p y$ only varies significantly for $y>15$, the wall-shear stress being constant (see figure \ref{fig:DR} (left)). We finally note that the production term 5, $\averph{\averper{\omega_j \p w/\p x_j } \p \averper{U}/\p y}$, decreases substantially in the oscillating-wall case, while the production term 8 and the transport term 9, which only involve fluctuating quantities, are largely unaffected.

%-----------------------------------------------------
\subsection{Transient response of turbulent enstrophy}
\label{sec:transient}

\begin{figure}
  \centering
  \psfrag{t}{$t$}
  \psfrag{y}{}
  \includegraphics[width=\columnwidth]{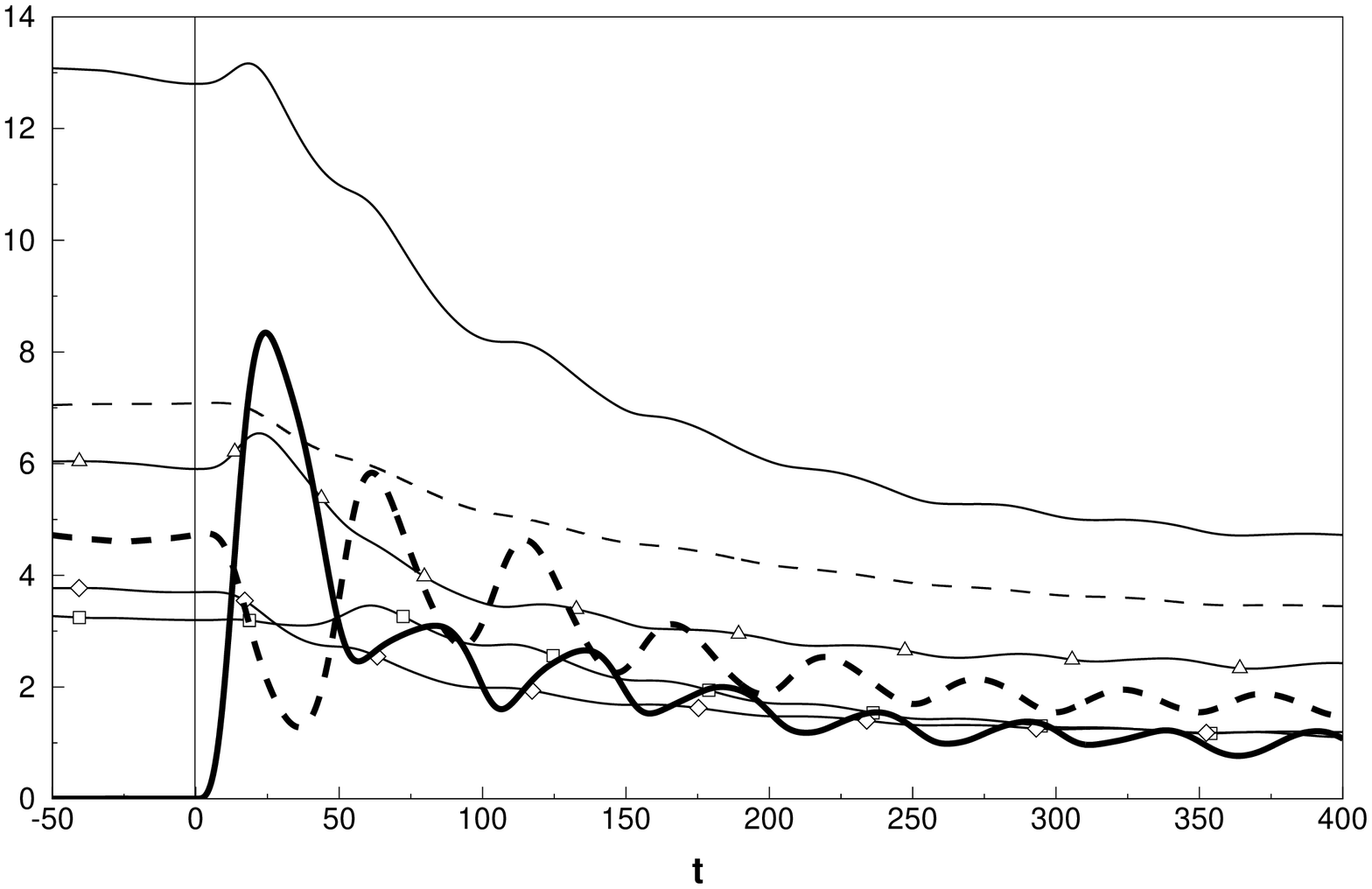}
  \caption{Temporal evolution of the space-averaged, $y$-integrated turbulent kinetic energy (thin dashed line), turbulent enstrophy (thin solid line), three vorticity terms forming the turbulent enstrophy ($\omega_x$ denoted by squares, $\omega_y$ denoted by diamonds, and $\omega_z$ denoted by triangles), and term 2 (thick dashed line) and term 3 (thick solid line) in (\ref{eq:enstrophy}). The oscillation starts at $t=0$. Terms 2 and 3 are multiplied by a scale factor of 30, while the kinetic energy is divided by a scale factor of 100.}
  \label{fig:transient}
\end{figure}
\begin{figure}
  \centering
  \psfrag{t}{$t$}
  \psfrag{y}{}
  \psfrag{b}{\tiny \mymktiny{2}}
  \psfrag{c}{\tiny \mymktiny{3}}
  \psfrag{d}{\tiny \mymktiny{4}}
  \psfrag{e}{\tiny \mymktiny{5}}
  \psfrag{f}{\tiny \mymktiny{6}}
  \psfrag{g}{\tiny \mymktiny{7}}
  \psfrag{h}{\tiny \mymktiny{8}}
  \psfrag{i}{\tiny \mymktiny{9}}
  \psfrag{j}{\tiny \mymktiny{10}}
  \psfrag{m}{\tiny \mymktiny{11}}
  \includegraphics[width=\columnwidth]{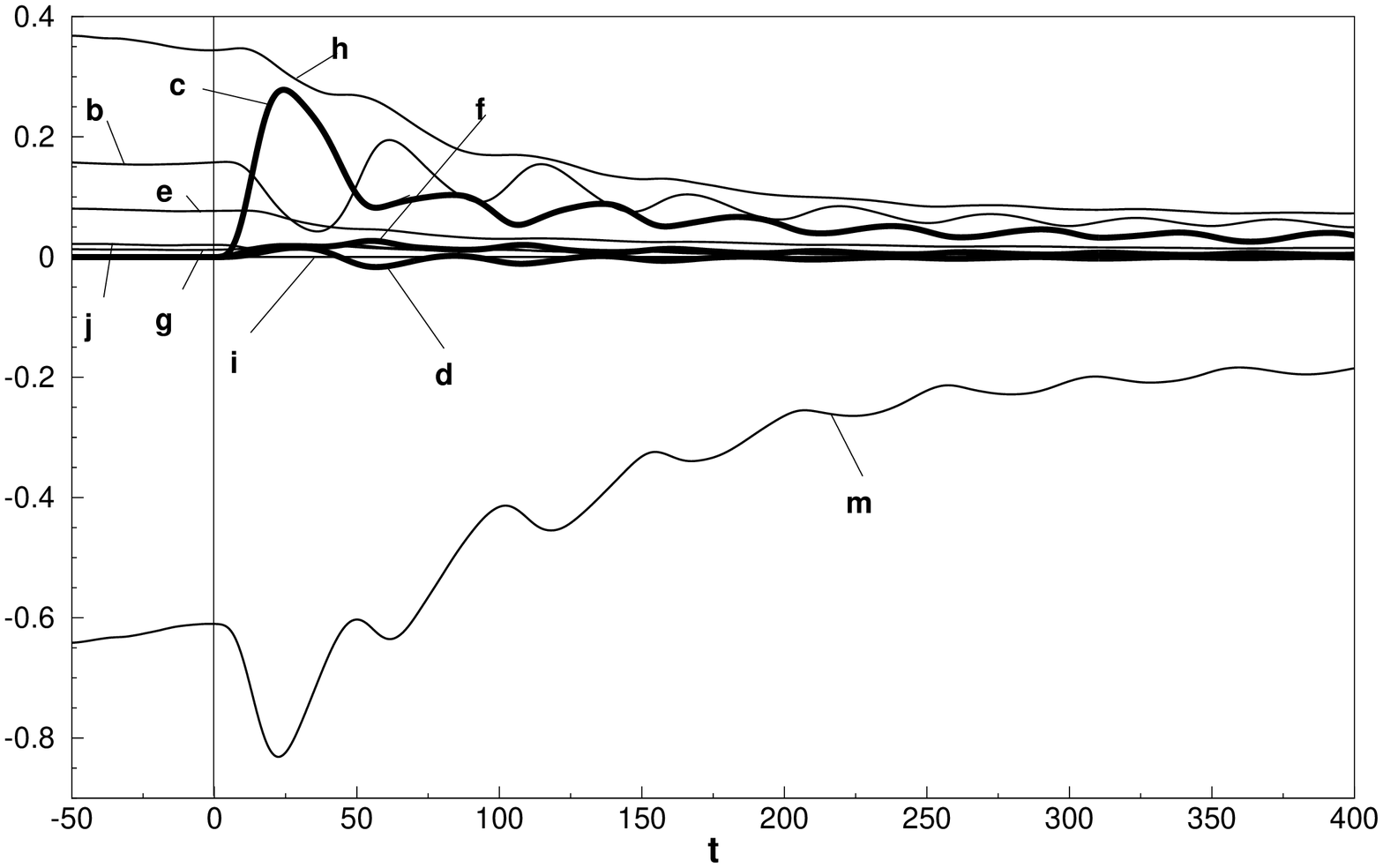}
  \caption{Temporal evolution of the space-averaged, $y$-integrated terms in the turbulent enstrophy equation (\ref{eq:enstrophy}). Thick lines highlight terms only occurring during the wall motion.}
  \label{fig:transient_enstrophy}
\end{figure}
Although it is clear from the discussed results that the spanwise shear layer $\averper{W}$ enhances the turbulent dissipation, it is difficult to verify the third scenario proposed in \S\ref{sec:total} through the energy balance shown in figure~\ref{fig:energy-balance}, i.e. that the relative contribution of $\mathcal{D}_\turb$ to the global balance decreases. Furthermore, by studying the global balance no information is gained on why the TKE decreases when the walls move. We therefore study the effect of the oscillation at very short times, in line with \cite{quadrio-ricco-2003} and \cite{xu-huang-2005}, who have used the same approach to study the Reynolds stress budget. 

The temporal evolution of the space-averaged, $y$-integrated turbulent kinetic energy, turbulent enstrophy, squares of the turbulent vorticity components, and terms 2 and 3 in (\ref{eq:enstrophy}) are shown in figure~\ref{fig:transient}, where the wall is in motion for $t>0$. On this short time scale, terms 2 and 3 show an oscillating behaviour, whose period matches well that of the wall forcing.
Upon the beginning of the oscillation, term 3, denoted by the thick solid line, grows abruptly until $t=25$, i.e. at a quarter of the oscillation period. It gives a transient production of turbulent enstrophy, and, specifically, of $\omega_z$, whose production is related to the spanwise tilting of $\omega_y$ by the Stokes layer, $\omega_y \p\averxz{W}/\p y$, in the transport equation of $\omega_z$. Term 2, which is indicated by the thick dashed line and is non-zero before the start of the oscillations, drops substantially up to $t$=30 and then becomes larger than term 3. As $\p \averxz{U}/\p y$ is unchanged on such a short time scale, the decrease of term 2 is directly linked to $\averxz{\omega_x\omega_y}$. This can be interpreted as the wall oscillation causing a change of the phase relationship between the low-speed streaks (related to $\omega_y$) and the quasi-streamwise vortices (related to $\omega_x$). This scenario is in line with early suggestions on the effects of the forcing on near-wall coherent structures \citep{baron-quadrio-1996}. This mechanism is indirect, since $\averxz{W}$ does not appear explicitly in term 2. 
Term 2 also appears in the transport equation for $\omega^2_x$ and represents the tilting of $\omega_y$ along $x$ due to $\p \averxz{U}/\p y$. Figure~\ref{fig:transient} indeed shows that the first two instantaneous peaks of term 2 agree fairly well with those of the transient evolution of $\omega_x^2$.
The long-term behaviour of term 2, namely the slight increase and the outward shift of its maximum, is due to the increase of $\p \averxz{U}/\p y$ in the bulk of the channel, caused by the increase in mass flow rate. Figure~\ref{fig:transient_enstrophy} shows the temporal evolution of all the space-averaged, $y$-integrated terms in the enstrophy equation (\ref{eq:enstrophy}). Term 3 is the cause of the short-term transient changes in the enstrophy balance, which is evinced by the resemblance of its temporal history with the one of term 11, the turbulent enstrophy dissipation, up to $t$=50.

\begin{figure}
  \centering
  \psfrag{x}{$t$}
  \psfrag{y}{}
  \includegraphics[width=\columnwidth]{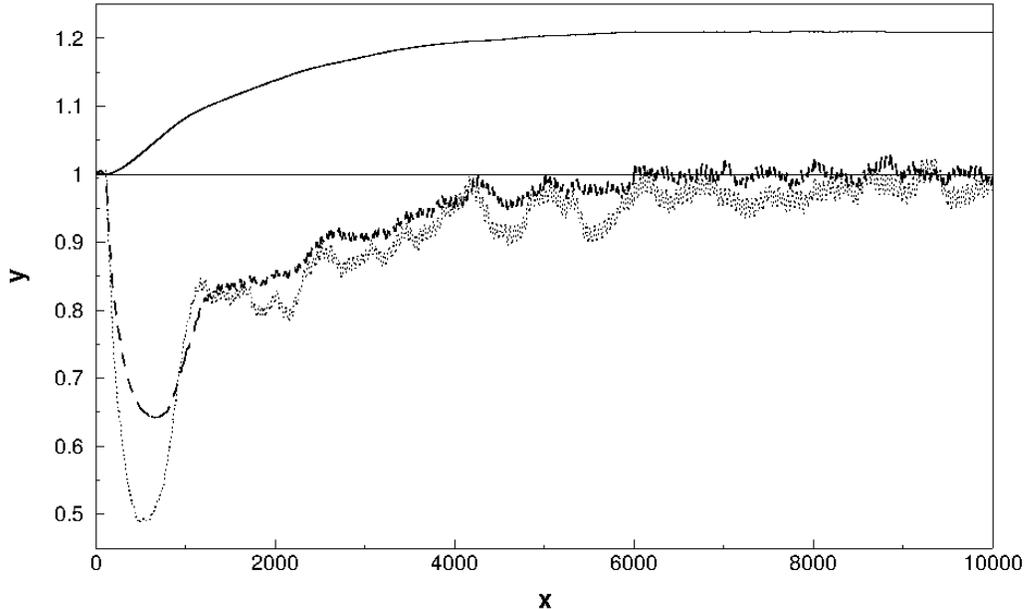}
  \caption{Temporal evolution of the space-averaged wall friction (dashed line), the TKE (dotted line), and the mass flow rate (solid line). The TKE follows the temporal behaviour of the friction, whereas it is graphically evident how the unbalanced friction causes a flow acceleration with a consequent increase of the mass flow rate. All values are normalized by the average uncontrolled case. 
 }
  \label{fig:longtransient}
\end{figure}

The turbulent dissipation is therefore enhanced, which causes the monotonic decrease of TKE. This feeds back onto the turbulent vorticity and onto term 3, which are both diminished because of the weakened turbulent activity. As a direct consequence of the attenuation of TKE, the streamwise mean flow accelerates, thereby increasing the streamwise mean velocity. This is evident from the mean streamwise momentum equation, 
\begin{equation}
\label{eq:mean-x-mom}
- \Pi = \frac{\p \averxz{U}}{\p t} - \frac{\p^2 \averxz{U}}{\p y^2} + \frac{\p \averxz{uv}}{\p y},
\end{equation}
because $-\Pi$ is constant (and positive) and the convective term $\p \averxz{U}/\p t$ must be positive to counteract the decay of the convective transport due to the Reynolds stresses (which are larger than the mean viscous terms $\p^2 \averxz{U}/\p y^2$). The TKE continuously decreases because, although the turbulent dissipation and production are both attenuated, the latter is proportionally smaller.  
As the streamwise flow accelerates, all the quantities decrease up to $t \approx 400$. This mirrors the transient behaviour of the turbulence under constant mass flow rate conditions, studied by \cite{quadrio-ricco-2003}: on such a short time scale, there is no difference between the two constraints. This is further supported by the initial attenuation of the wall-shear stress, $\p \averxz{U}/\p y|_{y=0}$, which is an immediate consequence of the acceleration of the mass flow rate. This is shown by integrating (\ref{eq:mean-x-mom}) along $y$,
\[
- \Pi h = \frac{\p}{\p t} \left( \int_0^h \averxz{U} \dif y \right) + \pipe{\frac{\p \averxz{U}}{\p y}}_{y=0}.
\]
As $-\Pi$ is positive and the flow-rate term on the r.h.s. is positive, the wall-shear stress must be smaller than its steady-state value during the transient evolution. The value of the wall stress eventually re-establishes itself in the new fully-developed regime to the value imposed by the constant $\Pi$. This is evident in figure \ref{fig:longtransient}, where the long-time evolution of TKE and of the mass flow rate are also shown.

The temporal evolution has helped to clarify the action of the wall motion on the turbulence dynamics. In  particular, as the oscillation initially enhances the turbulent dissipation, it is shown that the turbulent activity is suppressed due to the dissipative nature of the flow. In the new quasi-equilibrium regime reached after the long transient has elapsed, the flow therefore requires a relatively lower level of turbulent dissipation because TKE is lower. The third scenario discussed in \S\ref{sec:total} is therefore at work and term 3 is key to explaining the lower contribution of $\mathcal{D}_\turb$ in the global balance. 

\begin{figure}
\vspace{1cm}
  \centering
    \psfrag{S}{\small \bf Short}
    \psfrag{ss}{\small $t<50$}
    \psfrag{I}{\small \bf Intermediate}
    \psfrag{ii}{\small $50<t<400$}
    \psfrag{L}{\small \bf Long}
    \psfrag{ll}{\small \bf $t>400$}
    \psfrag{init}{\small Initial state}
    \psfrag{new}{\small `Drag reduction'}
    \psfrag{Ubup}{\small $\int_0^h \averxz{U} \dif y \uparrow$}
    \psfrag{dUdtup}{\small $\boxed{\frac{\p \averxz{U}}{\p t} > 0}$}
    \psfrag{reydown}{\small $\frac{\p \averxz{uv}}{\p y} \downarrow$}
    \psfrag{term-3}{\small $\averxz{\omega_z \omega_y} \frac{\p \averxz{W}}{\p y} \uparrow$}
    \psfrag{om2up}{\small $\averxz{\omega_i \omega_i} \uparrow$}
    \psfrag{DTup}{\small $\boxed{\mathcal{D}_\turb \uparrow}$}
    \psfrag{TKEdown}{\small TKE $\downarrow$}
  \includegraphics[width=\textwidth]{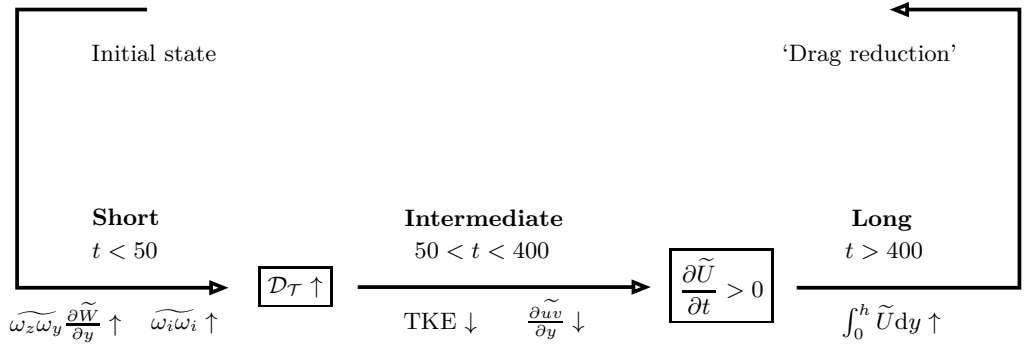}
  \caption{Schematic of the physical mechanism leading to skin-friction drag reduction by wall oscillations, as discussed in \S\ref{sec:transient}. The vertical arrows indicate whether the quantities increase or decrease.}
  \label{fig:physicsnew}
\end{figure}

We close this section with the schematic in figure \ref{fig:physicsnew}. The crucial physical processes during the temporal flow evolution from the start-up of the wall motion to the new fully-developed regime are shown. This last regime is indicated with `Drag reduction', although it should be recalled that, in the present context, the turbulent drag is eventually unvaried by design, and that the effect of the oscillations is to increase the mass flow rate. 

%------------------------------------------------------------------------------------------
\subsection{Physical interpretation of $\averph{\averper{\omega_z \omega_y} \frac{\p \averper{W}}{\p y}}$}
\label{sec:physics}

\begin{figure}
  \centering
  \psfrag{x}{$y$}
  \psfrag{y}{$\averper{\omega_z\omega_y}$}
  \psfrag{a}{$3a$}
  \psfrag{b}{$3b$}
  \psfrag{c}{$3c$}
  \psfrag{d}{$3d$}
  \includegraphics[width=\textwidth]{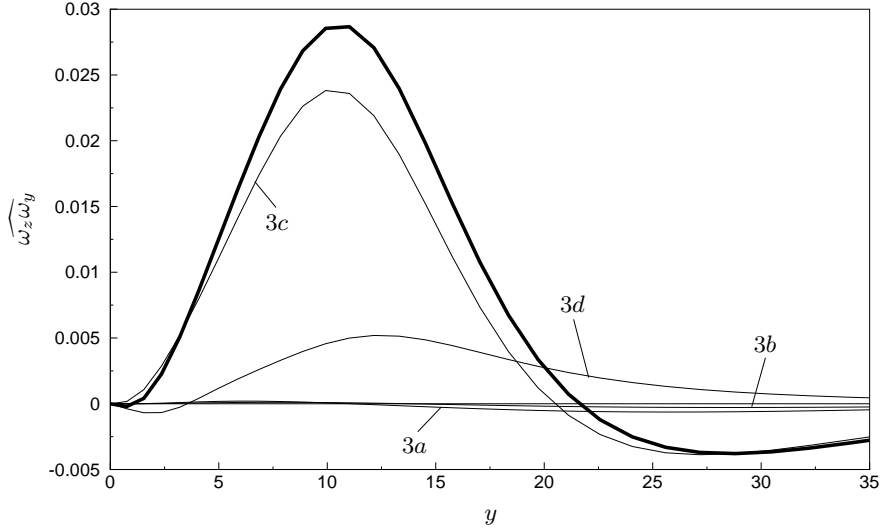}
  \caption{Wall-normal profile of $\averper{\omega_z\omega_y}$ at the phase at which it reaches its maximum (thick line). Thin lines indicate the sub-terms in \eqref{eq:vortproduct}.}
  \label{fig:dudz-dudy}
\end{figure}
Term 3 in the enstrophy equation \eqref{eq:enstrophy}, $\averph{\averper{\omega_z \omega_y} \p \averper{W} / \p y}$, has been found to be dominant and largely responsible for the change of global turbulent enstrophy and thus for drag reduction. It is positive and therefore indicates a production of turbulent vorticity, i.e. the mean spanwise flow shear $\p \averper{W} / \p y$ acts on the turbulence structures represented by the term $\averper{\omega_z\omega_y}$ to increase the turbulent enstrophy. We proceed to investigate the physical meaning of such interaction in more detail.

The quantity $\averper{\omega_z\omega_y}$ can be expanded as:
\begin{equation}
 \averper{\omega_z \omega_y} 
 =
   \underbrace{\averper{\frac{\p v}{\p x} \frac{\p u}{\p z}}}_{3a}
 - \underbrace{\averper{\frac{\p v}{\p x} \frac{\p w}{\p x}}}_{3b}
 - \underbrace{\averper{\frac{\p u}{\p y} \frac{\p u}{\p z}}}_{3c}
 + \underbrace{\averper{\frac{\p u}{\p y} \frac{\p w}{\p x}}}_{3d}.
\label{eq:vortproduct}
\end{equation}
Figure \ref{fig:dudz-dudy} shows that $3c$ is the largest contributor to $\averper{\omega_z \omega_y}$, and that the next largest sub-term in magnitude is term $3d$. This confirms the order-of-magnitude analysis in the Appendix. Terms $\p u / \p z$ and $\p u / \p y$, contributing to term 3c, may be linked separately to the dynamics of the turbulent low-speed streaks. In the near-wall region, $\p u / \p z$ marks the lateral flanks of the low-speed streaks, i.e. the interfaces of the low-velocity and high-velocity regions, while $\p u / \p y$ is related to the eruption of near-wall low-speed fluid to higher locations and to the sweep-like motion of high-speed fluid toward the wall. It is also noted that the peak location of $3c$ matches well that of $\mathcal{D}_\turb$ in figure~\ref{fig:budget_tke}, suggesting that the enhancement of $\mathcal{D}_\turb$ is connected to term $\averph{\averper{\omega_z \omega_y} \p \averper{W} / \p y}$. 

\begin{figure}
\psfrag{x}{$x$}
\psfrag{z}{$z$}
  \centering
  \includegraphics[width=0.49\columnwidth]{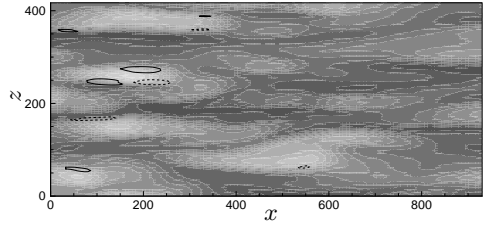}
  \caption{Instantaneous contour plots of $u$ and $\omega_y \omega_z$, for the fixed-wall case, in the $x-z$ plane at $y=6$. For the sake of clarity, only a fraction of the computational domain is shown. Gray shading represents the level of streamwise velocity fluctuations, with white corresponding to the maximum values and black to the minimum values; max$|u|=8$. Contour lines represent values of the quantity $\omega_y \omega_z$; contour levels start from $\pm 0.125$, are spaced by $0.25$, and negative values are dashed.}
  \label{fig:contour-ref}
\end{figure}
\begin{figure}
\psfrag{x}{$x$}
\psfrag{z}{$z$}
  \centering
  \includegraphics[width=0.49\columnwidth]{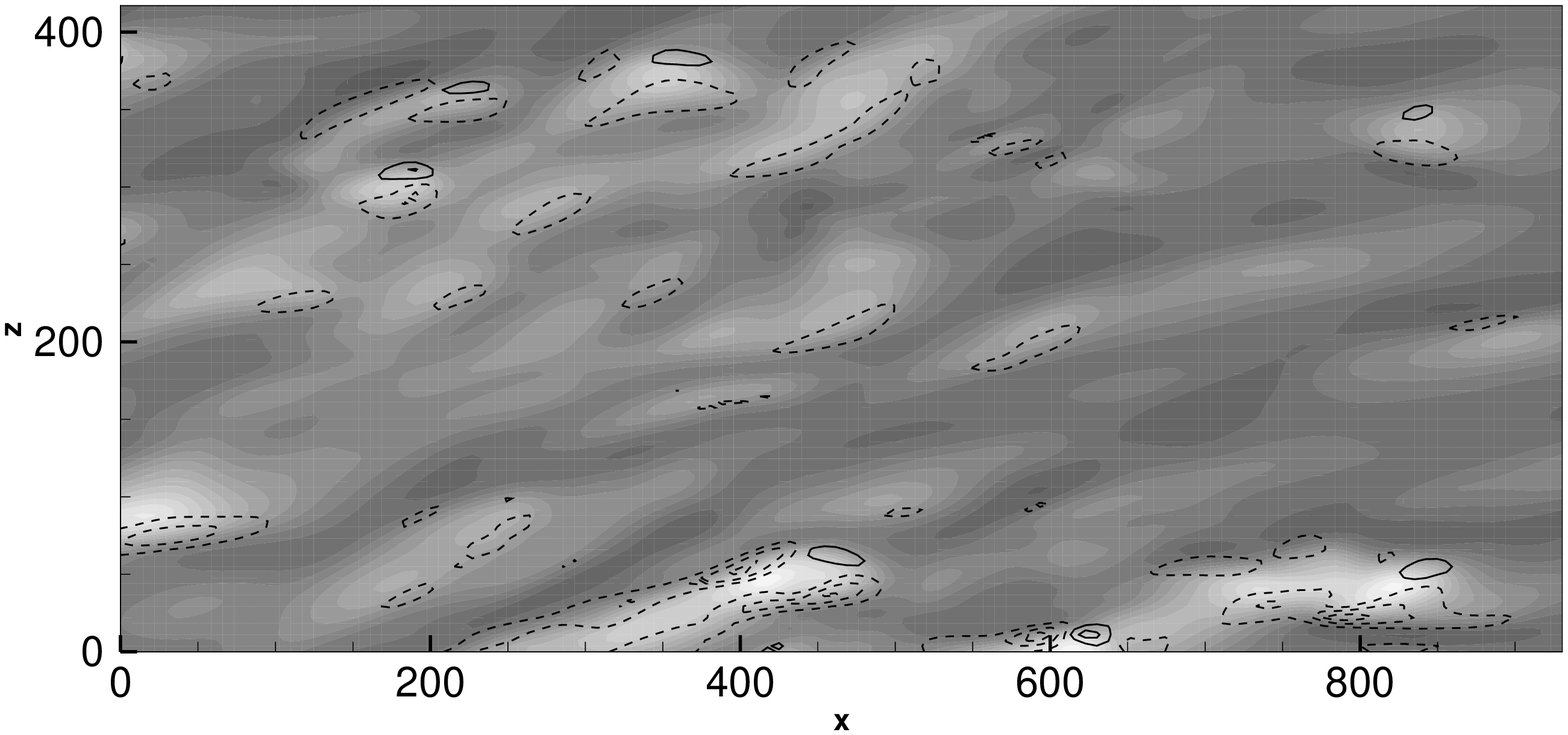}
  \includegraphics[width=0.49\columnwidth]{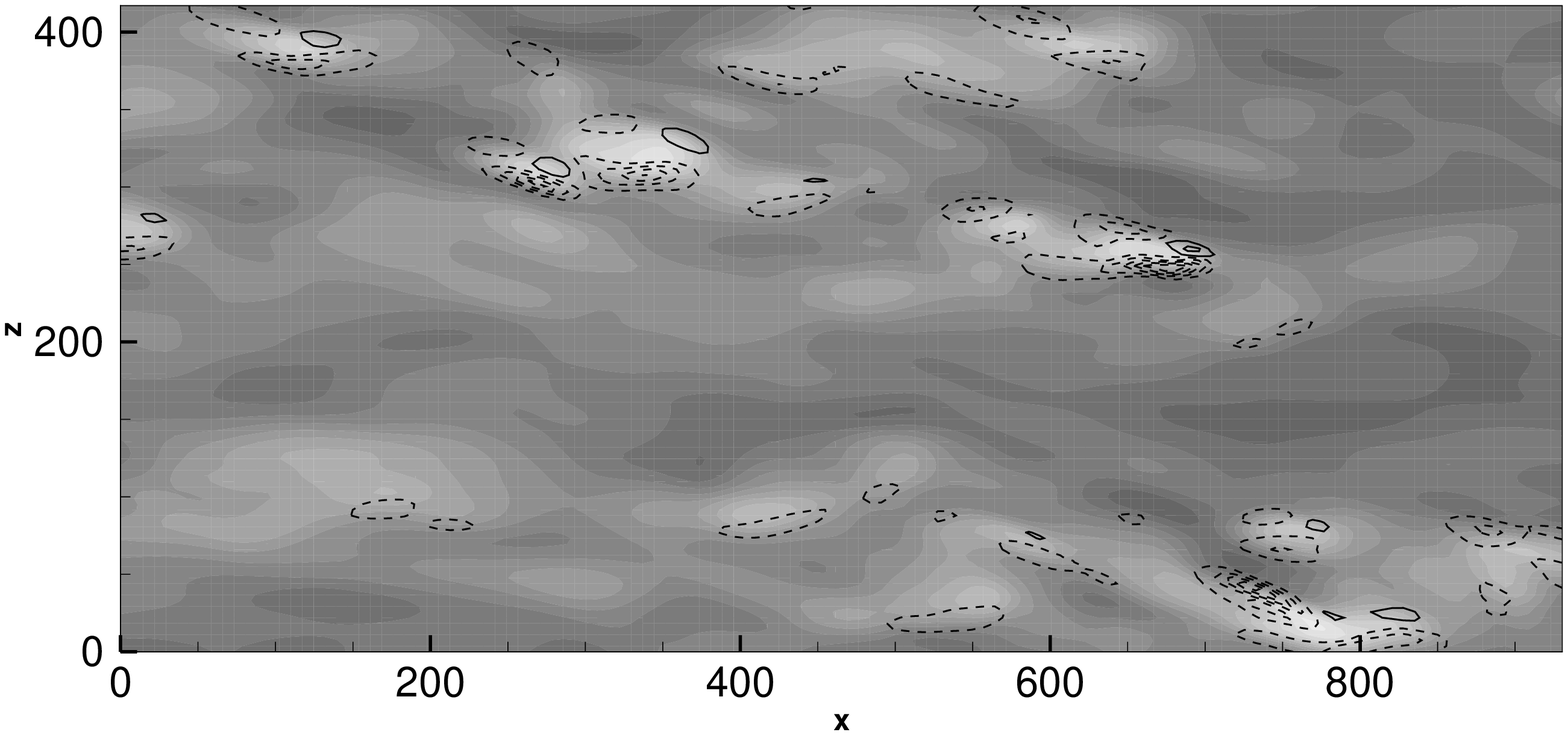} \\
  \includegraphics[width=0.49\columnwidth]{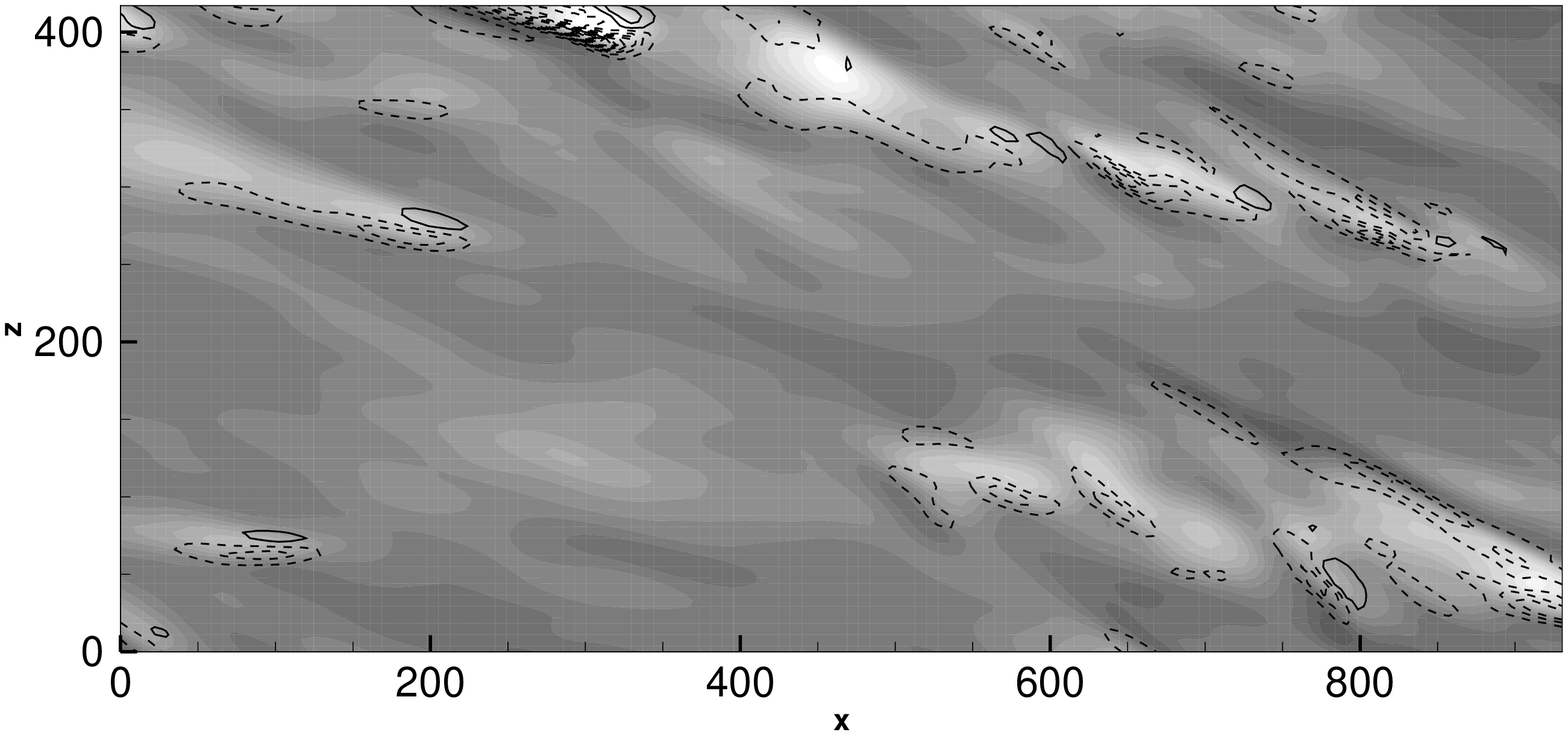}
  \includegraphics[width=0.49\columnwidth]{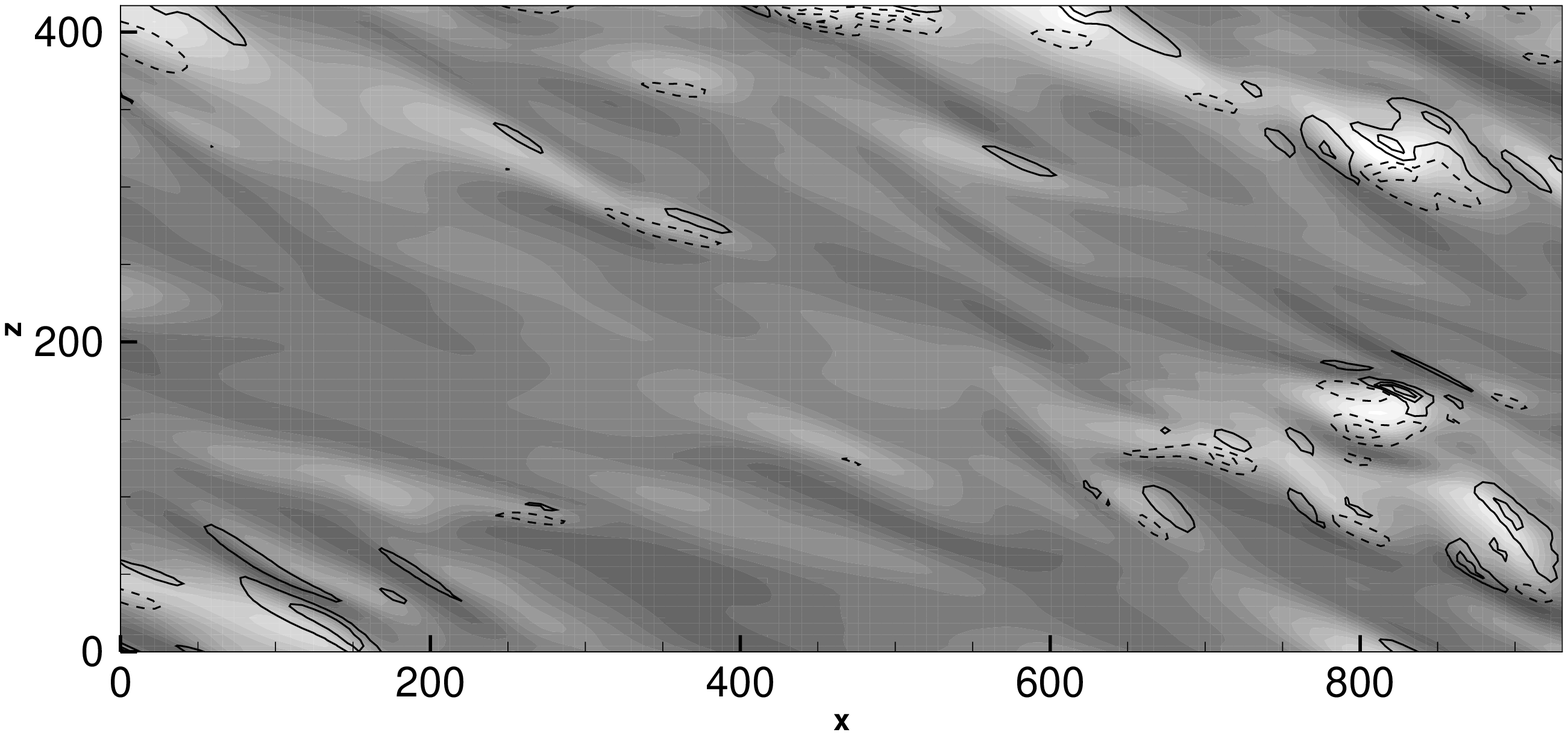}
  \caption{Instantaneous contour plots of $u$ and $\omega_y \omega_z$, for the oscillating-wall case, in the $x-z$ plane at $y=6$. From left to right and top to bottom: $\phi$=$0, \pi/4, \pi/2, 3\pi/4$. Legend is the same of figure \ref{fig:contour-ref}.}
  \label{fig:contour-ow}
\end{figure}
%Here above cuts are computed from fields 33-34-35-36 of caseT100-160 with the code fld2dat.cpl

Figure \ref{fig:contour-ref} shows contour plots of streamwise velocity fluctuations and $\omega_y \omega_z$ in the $x-z$ plane at $y=6$ for the fixed-wall case. Low- and high-speed streaks show the characteristic streamwise-stretched shape. Low-speed streaks are longer and thinner than the high-speed ones. Regions of high magnitude of $\omega_y \omega_z$ are concentrated near the wall. At $y=6$, they appear sporadically and always occur at the sides of high-speed regions. Figure \ref{fig:contour-ow} shows the contours plots for the oscillating-wall case at four different phases, where the characteristic cyclic tilting of the near-wall structures is evident \citep{quadrio-ricco-2003}. The streaks are less energetic, which confirms the attenuation of $\averph{\averper{u^2}}$, shown in figure \ref{fig:rms}. Regions of high $|\omega_y \omega_z|$ show an analogous tilting, owing to their relationship with the velocity streaks. The number, the amplitude, and the spatial size of the $\omega_y \omega_z$ pockets strongly increase during the wall motion, in line with the observed intensified enstrophy fluctuations. 

The interaction between the large-scale oscillating shear layer and the underlining vortical structures can be modeled by analogy with the rapid distortion theory problem of a large eddy stretching a smaller blob of vorticity, presented by \cite{davidson-2004} at page 213. We consider small-scale vorticity structures being stretched and compressed by the large-scale action of the spanwise layer and we extend the model by \cite{davidson-2004} to include the viscous dissipation effects. It is assumed that the forcing induced by the wall motion is more energetic than the turbulent fluctuations and it operates on a longer time scale. To simplify the problem further, the focus is on the dynamics at a wall-normal location $y \approx 6$, where, as shown in figure \ref{fig:budget-enstrophy}, the vorticity production given by the spanwise layer, i.e. term 3 in (\ref{eq:enstrophy}), is largely balanced by the viscous enstrophy dissipation, i.e. term 11 in (\ref{eq:enstrophy}). The gradient of the spanwise layer, indicated by $G$, is taken as constant in the small region considered.

In the $y-z$ plane, the enstrophy dynamics may thus be distilled into the following simplified equation
\begin{equation}
\label{eq:enstrophy-simplified}
\frac{1}{2} \frac{\p }{\p t}\left(\omega_y^2+\omega_z^2\right) = \omega_z\omega_y G - \left(\frac{\p \omega_y}{\p y}\right)^2 - \left(\frac{\p \omega_z}{\p y}\right)^2,
\end{equation}
where only the terms involving wall-normal gradients are retained amongst the dissipation terms because they are dominant as revealed by the order-of-magnitude analysis in the Appendix. The terms on the r.h.s. of (\ref{eq:enstrophy-simplified}) may be written in matrix form as follows
\[
\omega_z\omega_y G =
[\omega_y ; \omega_z]
\begin{bmatrix}
0 & G/2 \\
G/2 & 0
\end{bmatrix}
\begin{bmatrix}
\omega_y  \\
\omega_z
\end{bmatrix},
\]
\[
\left(\frac{\p \omega_y}{\p y}\right)^2 + \left(\frac{\p \omega_z}{\p y}\right)^2 =
\frac{\p}{\p y}[\omega_y ; \omega_z]
\begin{bmatrix}
1 & 0 \\
0 & 1
\end{bmatrix}
\frac{\p}{\p y}
\begin{bmatrix}
\omega_y  \\
\omega_z
\end{bmatrix}.
\]
\begin{figure}
  \centering
  \psfrag{y}{$y$}
  \psfrag{z}{$z$}
  \psfrag{n}{$x_n$}
  \psfrag{s}{$x_s$}
  \psfrag{a}{$\alpha$}
  \psfrag{o}{$\boldsymbol\omega_{yz}$}
  \includegraphics[width=0.5\columnwidth]{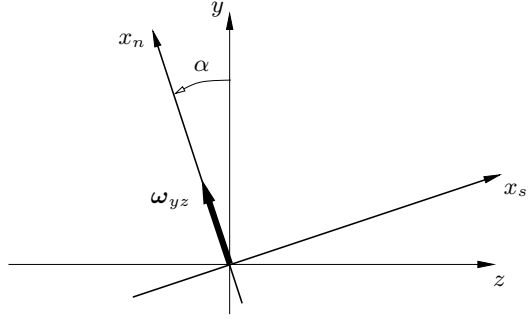}
  \caption{Coordinate systems in $y-z$ plane for turbulent vorticity vector $\boldsymbol\omega_{yz}$.}
  \label{fig:rotation}
\end{figure}
As shown in figure \ref{fig:rotation}, a set of perpendicular axes $(x_n,x_s)$ may be considered where $x_n$ is orientated along the vorticity vector in the $y-z$ plane, $\boldsymbol\omega_{yz}=[\omega_y ; \omega_z]$. The angle $\alpha$ is defined between $\boldsymbol\omega_{yz}$ and the wall-normal axis $y$. In the new set of coordinates, the terms on the r.h.s. of (\ref{eq:enstrophy-simplified}) are written as
\[
\omega_z\omega_y G =
[\omega_n ; 0]
\begin{bmatrix}
S_{nn} & S_{ns} \\
S_{sn} & S_{ss}
\end{bmatrix}
\begin{bmatrix}
\omega_n  \\
0
\end{bmatrix}= S_{nn} \omega_n^2,
\]
\[
\left(\frac{\p \omega_y}{\p y}\right)^2 + \left(\frac{\p \omega_z}{\p y}\right)^2 =
\frac{\p}{\p y}[\omega_n ; 0]
\begin{bmatrix}
D_{nn} & D_{ns} \\
D_{sn} & D_{ss}
\end{bmatrix}
\frac{\p}{\p y}
\begin{bmatrix}
\omega_n  \\
0
\end{bmatrix}= D_{nn} \left(\frac{\p \omega_n}{\p y}\right)^2,
\]
where $\omega_n^2 = \omega_y^2 + \omega_z^2$, $S_{nn} = \sin \alpha \cos \alpha G$, and $D_{nn}=1$, $D_{ns}=0$. The other components of the strain-rate tensor, $S_{ns},S_{sn}$, only contribute to change the direction of the vorticity vector, not its magnitude. Equation (\ref{eq:enstrophy-simplified}) is written as
\begin{equation}
\label{eq:enstrophy-simplified-rotated}
\frac{1}{2} \frac{\p \omega_n^2}{\p t} = S_{nn} \omega_n^2  - \left(\frac{\p \omega_n}{\p y}\right)^2.
\end{equation}
Equation (\ref{eq:enstrophy-simplified-rotated}) may be integrated by Charpit's method \citep{garabedian-1964} to give
\begin{equation}
\label{eq:enstrophy-growth}
\omega_n = \omega_{n,0} \underbrace{\e^{\sin \alpha \cos \alpha G t}}_{\tiny \mbox{stretching}}
\underbrace{\e^{- \beta^2 t} \e^{- \beta y}}_{\tiny \mbox{dissipation}},
\end{equation}
where $\omega_{n,0}$ is the initial magnitude. The constant
\[
\beta = \frac{\p \omega_n/\p t}{\p \omega_n/\p y} \sim \frac{\lambda_y}{\lambda_t},
\]
where $\lambda_y$ represents the dissipative scale along the wall-normal direction, and $\lambda_t$ indicates the time scale of the turbulent fluctuations. 

Equation (\ref{eq:enstrophy-growth}) shows that the spanwise layer may stretch or compress the turbulent fluctuations depending on the sign of its gradient and the orientation of the vorticity vector. The spanwise layer works by stretching when $\sin \alpha \cos \alpha G>0$. Its action is null when the vorticity vector is parallel or perpendicular to the wall, and maximum when either i) $G$ is at its negative peak at that $y$ location and simultaneously $\boldsymbol\omega_{yz}$ is oriented at $\pi/4$ with respect to the axes and along the first or third quadrant in figure \ref{fig:rotation}, or ii) $G$ is at its positive peak and simultaneously $\boldsymbol\omega_{yz}$ is orientated at $\pi/4$ with respect to the axes and along the second or fourth quadrant in figure \ref{fig:rotation}.
It is further noted that the temporal rate of growth or decay of enstrophy is never larger than $G$, and that the exponential attenuation through the viscous effects is more intense in time than space. 

We conclude that turbulent vorticity is produced when $\sin \alpha \cos \alpha G > \beta^2$, i.e. when the shear-layer production, ruled by the intensity of the large-scale spanwise shear layer and by the orientation of the turbulent vorticity vector, overcomes the viscous dissipation, whose dynamics is linked to the time scale and the wall-normal spatial scale of the fluctuating vorticity.

%------------------------------------------------------------
\subsection{Drag reduction and production of turbulent enstrophy}
\label{sec:scaling}

\begin{figure}
  \centering
  \psfrag{x}{$\averglob{\averper{\omega_z \omega_y} \p \averper{W} / \p y}$}
  \psfrag{y}{$R$}
  \psfrag{a}{$T$=$8$}
  \psfrag{b}{$T$=$21$}
  \psfrag{c}{$T$=$42$}
  \psfrag{d}{$T$=$100$}
  \includegraphics[width=\textwidth]{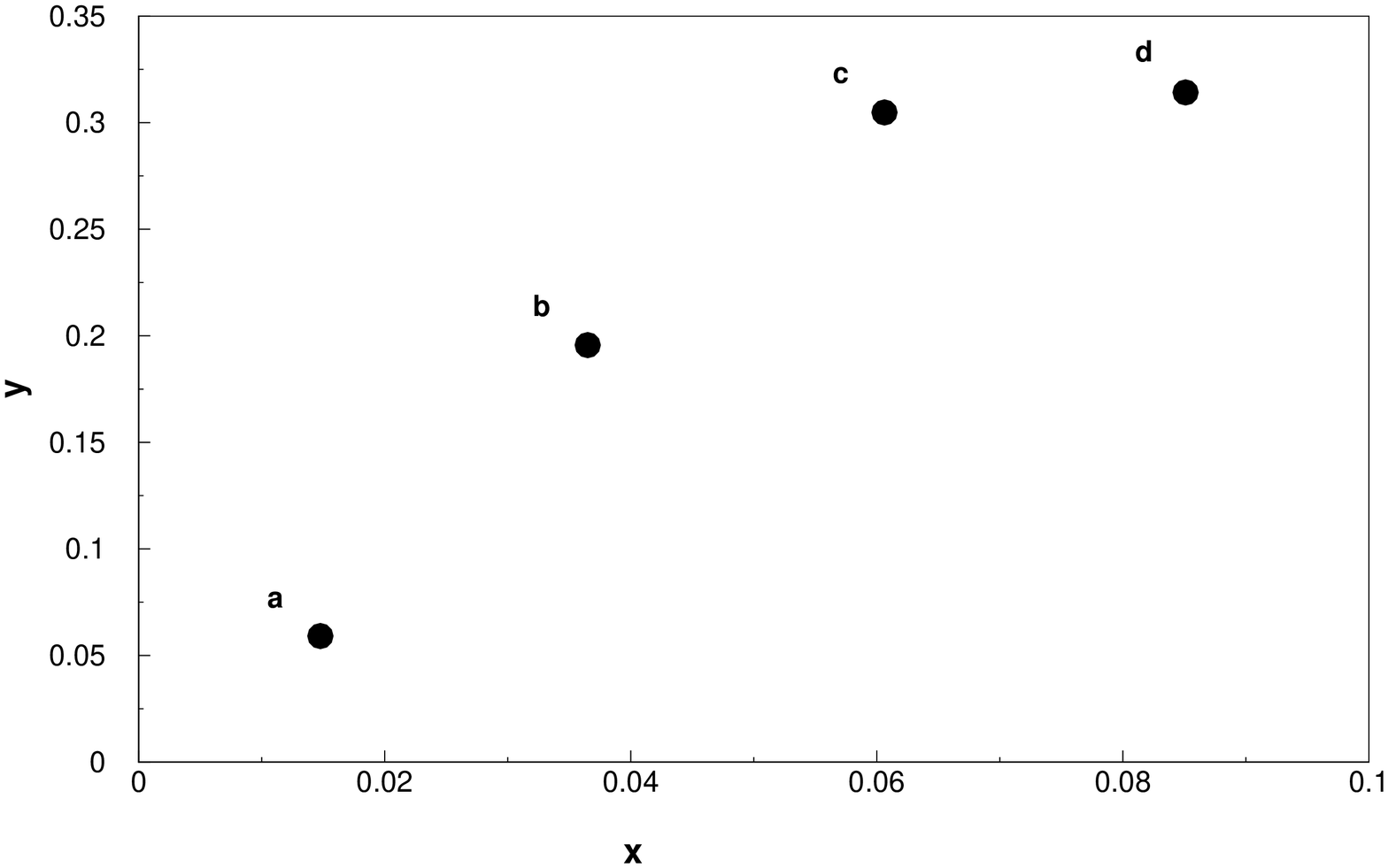}
  \caption{Drag reduction as function of $\averglob{\averper{\omega_z \omega_y} \p \averper{W} / \p y}$ at different oscillations periods. For small values of $T$ the two quantities are linearly related.}
  \label{fig:DRvsTHREE}
\end{figure}

The importance of the enstrophy production term 3 has been revealed. However, our conclusions are only based on one flow condition, $A=12$ and $T=100$, yielding $R=0.31$. It remains to show whether this result can be generalized to other periods of oscillation.
To this purpose, additional simulations have been carried out by changing $T$ and leaving all the other parameters unvaried. Figure \ref{fig:DRvsTHREE} shows that $R$ is linearly proportional to $\averglob{\averper{\omega_z \omega_y} \p \averper{W} / \p y}$ up to the conditions where $R \approx 0.30$, which corresponds to $T = 42$. The linearity between $R$ and the global value of enstrophy production begins to lose its validity at $T=T_{opt}$.

The global value of the enstrophy term is shown because $R$ is linked via \eqref{eq:cf-ub} to the change of $U_b$, a global quantity as defined in \eqref{eq:ub}. It is known that when $T$ is larger than the characteristic life time of the near-wall turbulent structures \citep{quadrio-ricco-2004}, drag reduction drops as the forcing is slow enough to decouple from the near-wall turbulence dynamics. At high $T$, the oscillating wall is not expected to induce drag reduction because the near-wall structures are too slowly affected and thus tend to re-establish their natural dynamics between sweeps of the Stokes layer. Indeed, the case at $T=500$ (not shown in the figure) gives a small $R=0.06$, and the linear relationship with the global term 3 is lost. On the other extreme, the wall motion becomes ineffective and produces small $R$ at small $T$ owing to the limited wall-normal extent of the oscillating Stokes layer.

%------------------------------------------------------------------------
\subsection{A note on the scaling parameter by \cite{quadrio-ricco-2004}}
\label{sec:turbulent-omega-scaling}

In previous works, attempts have been made to identify a forcing parameter which scales linearly with drag reduction. 
In particular, \cite{choi-xu-sung-2002} have introduced a parameter that has been shown by \cite{quadrio-ricco-2004} to relate linearly to the amount of skin-friction reduction for periods of oscillations smaller than or comparable with the optimum. (At larger periods, the above-mentioned decoupling between the forcing and the near-wall turbulence takes place.) This scaling parameter was constructed by combining a characteristic length scale related to the wall-normal distance at which the wall motion affects the turbulent structures and the maximum spanwise acceleration of the Stokes layer. It reads:
\[
\mathcal{S} = 2 \sqrt{\frac{\pi}{T}} \ln \left( \frac{A}{A_{th}} \right) \exp \left( - \overline y \sqrt{\frac{2 \pi}{T}} \right),
\]
where $A_{th} \approx 1.2$ is a threshold velocity and $\overline y \approx 6.3$ is a wall-normal distance representative of the diffusion of the Stokes-layer viscous effects. 

It is then natural to inquire whether there exists a relationship between $\mathcal{S}$ and the global enstrophy production term (which also relates linearly to $R$, as shown in figure \ref{fig:DRvsTHREE}), here expressed as a new parameter,
\begin{equation}
\mathcal{S}_n =  \mathcal{S}_n(T,A) = \averglob{\averper{\omega_z \omega_y} \frac{\p \averper{W}}{\p y}}.
\label{eq:Sn}
\end{equation}
The first observation is that $\mathcal{S}$ can be written as
\begin{equation}
\label{eq:S-vort}
\mathcal{S} = \frac{2}{A} \ln \left( \frac{A}{A_{th}} \right) {\Omega_x}_m (\overline y),
\end{equation}
where ${\Omega_x}_m$ is the maximum streamwise vorticity of the Stokes layer at $y=\overline y$. Relation \eqref{eq:S-vort} endows $\mathcal{S}$ with a more direct and physically relevant meaning as it simply states that the drag reduction is linearly proportional to the maximum spanwise shear induced by the Stokes layer at constant $A$, and that such shear is most effective when at work at $y \approx 6.3$.
The fact that $\mathcal{S}$ relates well with drag reduction is not surprising in view of the scaling analysis based on the enstrophy production. The parameter $\mathcal{S}$ in \eqref{eq:S-vort} can be seen as a simplified version of $\mathcal{S}_n$ in \eqref{eq:Sn}. Although $\mathcal{S}_n$ is more elaborate as it possesses a precise physical meaning and involves averaging and wall-normal integration, the spanwise shear plays a key role in both expressions. The other point of note is that $\averph{\averper{\omega_z \omega_y} \p \averper{W}/\p y}$ reaches its maximum at $y \approx 6.5$ for optimum conditions of drag reduction, i.e. almost at the same distance at which the correlation between $R$ and $\mathcal{S}$ in \eqref{eq:S-vort} is maximum.

It is noted that other semi-empirical formulas linking $R$ and the wall oscillation parameters have been put forward. \cite{bandyopadhyay-2006} developed a formula also based on the effect of the Stokes layer on the near-wall turbulence. The central idea is that the Stokes layer cyclically re-orients the near-wall vorticity and the drag reduction is linearly related to the sine of the maximum angle of vorticity re-alignment with respect to the streamwise direction. This approach can be said to belong to the same family of ours, although the analysis in the present paper is based on the modification of the turbulent flow statistics (specifically related to the dissipation), while Bandyopadhyay's physical model is more directly inspired by the instantaneous action of the spanwise layer on the vortical coherent structures.

%-------------------------------
\section{Summary}
\label{sec:summary}

We have described via a DNS study how harmonic wall oscillations are capable of increasing the mass flow rate in a turbulent plane channel flow driven by a constant pressure gradient. The uniquely defined inner scaling brought about by the constant pressure gradient is exploited to ascertain how the oscillations modify the turbulence statistics. By looking at the energy fluxes in global form, it has emerged that the energy spent to drive the wall motion almost coincides with the viscous dissipation due to the oscillating spanwise layer, the difference taking the form of a small turbulence kinetic energy production term. The energy balance shows that the enhancement of energy intake due to the increased flow rate is mainly balanced by the combined increment of dissipation associated with the mean streamwise velocity profile and the turbulent dissipation. It is also revealed that the relative contribution of the latter to the total dissipation becomes smaller when the wall moves.

The spanwise oscillating layer shows a direct effect on the turbulent dissipation, which is conveniently expressed as the volume integral of the turbulent enstrophy. The amount of drag reduction relates linearly with the volume integral of an enstrophy production term induced by the spanwise shear layer. The study of the turbulent enstrophy transport equation reveals that this dominating enstrophy production term synthesizes the stretching of the vorticity lines by the oscillating layer, and therefore enhances the turbulent dissipation. The analysis of the short-term evolution of the flow after the beginning of the wall motion shows that the dissipative nature of the near-wall field is responsible for the attenuation of the turbulence intensity and thus for the increment of the bulk velocity. 

The study of the turbulent enstrophy in fully-developed conditions to evince which term dominates the physics and the analysis of the flow temporal evolution to discern how the new regime ensues can be both useful to investigate other turbulent flows modified by external agents, such as boundary layers affected by large-scale Lorentz or Coriolis forces, by wall transpiration, or by large temperature gradients. Furthermore, it would be of interest to use the approach based on the turbulent enstrophy to investigate the traveling-wave flow proposed by \cite{quadrio-ricco-viotti-2009} in drag-reduction and drag-increase conditions. 

%-------------------------------
\section*{Appendix. Order-of-magnitude analysis on turbulent enstrophy and dissipation equations}
\label{sec:order}

The order of magnitude of the terms arising because of the wall motion in the turbulent enstrophy equation (\ref{eq:enstrophy}) and in the turbulent dissipation equation can be estimated through an analysis similar to the one carried out by \cite{tennekes-lumley-1972} at pages 89 and 90. 
Two symbols are adopted, following the introductory discussion on the use of symbols in \cite{tennekes-lumley-1972}. The symbol $\sim$ denotes a crude approximation; it highlights the dependence of the term under scrutiny on the characteristic length and velocity scales of the turbulent motion. Upon decomposing an enstrophy term into sub-terms containing the fluctuating velocity components, the symbol $\mathcal{O}$ denotes its magnitude in terms of the dominant sub-term.
In \cite{tennekes-lumley-1972}, a generic length scale is assumed to describe the mean flow motion and the Taylor microscale is taken as the reference length scale for the turbulent fluctuations in all directions, suggesting that such an analysis is useful primarily for homogeneous isotropic turbulence. However, our interest is on the wall-bounded turbulence dynamics with wall oscillations, which is strongly anisotropic. It is therefore necessary to distinguish different length and velocity scales along the three Cartesian directions. 

The near-wall turbulent dynamics is characterized by three distinct length scales. The length scale of the disturbance along $z$ can be taken as $\lambda_z=\mathcal{O}(100)$, namely the characteristic spacing of the low-speed streaky structures \citep{kline-etal-1967}. As shown by \cite{ricco-2004}, the streaks spacing increases by about a fifth when $R \approx 0.3$, so that the order-of-magnitude estimate is still valid. The streaks length, $\lambda_x=\mathcal{O}(1000)$ for fixed-wall conditions, is representative of the disturbance flow along $x$ \citep{kline-etal-1967}. \cite{ricco-2004} has shown that  $\lambda_x$ decreases by about a third when $R \approx 0.3$. The order of magnitude of $\lambda_x=\mathcal{O}(1000)$ is therefore applicable under wall-oscillation conditions.
The length scale along $y$ for the mean flow is the spanwise boundary layer thickness $\delta$, defined here as the distance from the wall where the maximum $\averper{W}$ equals $\exp(-1)A$. As amply verified (see \cite{choi-xu-sung-2002}, amongst many), $\averper{W}$ agrees well with the laminar solution of the second Stokes problem for the flow induced by wall oscillations beneath a still fluid \citep{batchelor-1967}, so that the spanwise boundary layer thickness can be approximated well by $\delta=\sqrt{T/\pi}$. For $T=100$, $\delta \approx 5.7$, so that it can be assumed that $\delta=\mathcal{O}(10)$. The boundary layer thickness $\delta$ can be taken as the characteristic length scale for the near-wall disturbance flow because the oscillating boundary layer affects the turbulence in a region close to the wall whose width is comparable with $\delta$. This is shown in figure \ref{fig:rms} by the $\averph{\averper{u v}}$ profile being markedly affected only for $y < 25$ and by the $\averper{v w}$ profile  reaching its maximum at $y \approx 15$.
The mean-flow length scale becomes the length scale of the fluctuations along the direction of the shear also in other shear-driven phenomena, such as the penetration of free-stream turbulence into the Blasius boundary layer to form the Klebanoff modes \citep{leib-wundrow-goldstein-1999}. In that case, the wall-normal scale of the fluctuations within the boundary layer is the Blasius boundary layer thickness. The characteristic length scales along the Cartesian directions can therefore be taken as $\lambda_x > \lambda_z > \delta$.
As for the order of magnitude of the velocity components near the wall, as outlined by \cite{pope-2000} at page 283, both $u$ and $w$ show a linear growth near the wall, but the coefficient is larger for the streamwise component. The wall-normal component $v$ is smaller than both $u$ and $w$ because it grows quadratically from the wall. The hypothesis $u>w>v$ can therefore be adopted. The terminologies `larger' and `smaller' are used in the order-of-magnitude sense and the time-averaging symbol is omitted for brevity. 

The turbulent enstrophy equation (\ref{eq:enstrophy}) is considered first. The order of magnitude of terms 3, 4 and 6, arising in (\ref{eq:enstrophy}) because of the wall oscillation, is estimated. Term 3 can be first decomposed as follows.
\begin{equation}
 \mbox{Term} \ 3: \quad \averper{\omega_z\omega_y} \frac{\p \averper{W}}{\p y}
 = \left(
   \underbrace{\averper{\frac{\p v}{\p x}\frac{\p u}{\p z}}}_{3a}
 - \underbrace{\averper{\frac{\p v}{\p x}\frac{\p w}{\p x}}}_{3b}
 - \underbrace{\averper{\frac{\p u}{\p y}\frac{\p u}{\p z}}}_{3c}
 + \underbrace{\averper{\frac{\p u}{\p y}\frac{\p w}{\p x}}}_{3d}
 \right) \frac{\p \averper{W}}{\p y},
\label{eq:term3xx}
\end{equation}
and the order of magnitude of each sub-term is
\[
 3a \sim \frac{uv}{\lambda_x\lambda_z}, 
 3b \sim \frac{vw}{\lambda_x^2}, 
 3c \sim \frac{u^2}{\delta\lambda_z},  
 3d \sim \frac{uw}{\delta\lambda_x}, 
 \frac{\p \averper{W}}{\p y} \sim \frac{A}{\delta}.
\]
It is evident that term $3c$, $\averper{(\p u/\p y) (\p u/\p z)} \p \averper{W}/\p y$, is dominant. It follows that
\[
 \mbox{Term} \ {3}:\averper{\omega_z\omega_y}\frac{\p \averper{W}}{\p y} = \mathcal{O}\left(\frac{u^2 A}{\delta^2\lambda_z}\right).
\]
It further occurs that term $3d >$ term $3a >$ term $3b$. The magnitude of term 4 is estimated as follows.
\[
\begin{split}
\mbox{Term} \ 4: \quad \averper{\omega_i\frac{\p u}{\p x_i}}\frac{\p \averper{W}}{\p y} &= \left[\averper{\left(\frac{\p w}{\p y} - \frac{\p v}{\p z}\right)\frac{\p u}{\p x}} + \averper{\left(\frac{\p u}{\p z} - \frac{\p w}{\p x}\right)\frac{\p u}{\p y}} + \averper{\left(\frac{\p v}{\p x} - \frac{\p u}{\p y}\right)\frac{\p u}{\p z}}\right]\frac{\p \averper{W}}{\p y} \\
&=
\left(
 \underbrace{\averper{\frac{\p w}{\p y}\frac{\p u}{\p x}}}_{4a}
- \underbrace{\averper{\frac{\p v}{\p z}\frac{\p u}{\p x}}}_{4b}
- \underbrace{\averper{\frac{\p w}{\p x}\frac{\p u}{\p y}}}_{4c}
+ \underbrace{\averper{\frac{\p v}{\p x}\frac{\p u}{\p z}}}_{4d}
\right)
\frac{\p \averper{W}}{\p y},
\end{split}
\]
\[
 4a, 4c \sim \frac{uw}{\delta\lambda_x}, \
 4b, 4d \sim \frac{uv}{\lambda_x\lambda_z} 
\]
Terms $4a$ and $4c$ are larger than $4b$ and $4d$ because $w > v$ and $\delta < \lambda_z$, so that
\begin{equation}
\label{eq:term-4}
\mbox{Term} \ 4:  \averper{\omega_i\frac{\p u}{\p x_i}}\frac{\p \averper{W}}{\p y} = \mathcal{O}\left(\frac{uw A}{\delta^2\lambda_x}\right).
\end{equation}
Note that this represents an upper bound because terms $4a$ and $4c$ may add to produce a term of the order of magnitude given in (\ref{eq:term-4}) or give a term of smaller amplitude if these terms are of opposite sign. The magnitude of term $6$ can be estimated as follows.
\[
 \mbox{Term} \ 6: \quad -\averper{v\omega_x}\frac{\p^2 \averper{W}}{\p y^2} = \left(-\averper{v\frac{\p w}{\p y}} + \averper{v\frac{\p v}{\p z}}\right)\frac{\p^2 \averper{W}}{\p y^2},
\]
\[
 \averper{v\frac{\p w}{\p y}} \sim \frac{vw}{\delta}, \ 
 \averper{v\frac{\p v}{\p z}} \sim \frac{vv}{\lambda_z}, \
 \frac{\p^2 \averper{W}}{\p y^2} \sim \frac{A}{\delta^2}.
\]
The term $-\averper{v \p w/\p y}(\p^2 \averper{W}/\p y^2)$ is clearly dominant because $w>v$ and $\delta>\lambda_z$. It follows that
\[
\mbox{Term} \ 6: -\averper{v\omega_x}\frac{\p^2 \averper{W}}{\p y^2}=\mathcal{O}\left(\frac{vwA}{\delta^3}\right).
\]
In order to compare term $6$ with term $3$, we resort to the continuity equation, as follows
\[
\frac{\p v}{\p y} \sim \frac{\p u}{\p x} \Longrightarrow \frac{v}{\delta} \sim \frac{u}{\lambda_x},
\]
\[
\mbox{Term} \ 6: -\averper{v\omega_x}\frac{\p^2 \averper{W}}{\p y^2} = \mathcal{O}\left(\frac{vwA}{\delta^3}\right) = \mathcal{O}\left(\frac{uwA}{\delta^2 \lambda_x}\right).
\]
Since $u>w$ and $\lambda_x>\lambda_z$, one obtains
\[
 \mbox{Term} \ 3: \mathcal{O}\left(\frac{u^2 A}{\delta^2\lambda_z}\right) > \mbox{Term} \ 6: \mathcal{O}\left(\frac{uwA}{\delta^2 \lambda_x}\right).
\]
Terms $4$ and $6$ are either comparable, when the upper bound case for the order-of-magnitude estimate for term $4$ is considered, or term 4 $<$ term 6 if the two comparable leading terms in $4$ have opposite sign. It can be concluded that term 3 $>$ term 6 $\geq$ term 4, which is the result found through the numerical simulations.

The transport equation for the turbulent energy dissipation, called $\epsilon$ here
\[
  \epsilon \equiv \averper{\frac{\p u_i}{\p x_j}\left(\frac{\p u_i}{\p x_j}+\frac{\p u_j}{\p x_i}\right)},
\]
is now studied \citep{mansour-kim-moin-1989,fischer-jovanovic-durst-2001}. For the case of turbulent channel flow with spanwise wall oscillations, the equation reads
\begin{equation}
\label{eq:dissipation}
 \begin{split}
  \underbrace{\frac{1}{2}\frac{\p\epsilon}{\p \tau}}_{1} =& -
  \underbrace{\averper{\frac{\p u}{\p x_i}\frac{\p v}{\p x_i}}\frac{\p \averper{U}}{\p y}}_{2} -
  \underbrace{\averper{\frac{\p w}{\p x_i}\frac{\p v}{\p x_i}}\frac{\p \averper{W}}{\p y}}_{3} -
  \underbrace{\averper{\frac{\p u_i}{\p x}\frac{\p u_i}{\p y}}\frac{\p \averper{U}}{\p y}}_{4} -
  \underbrace{\averper{\frac{\p u_i}{\p z}\frac{\p u_i}{\p y}}\frac{\p \averper{W}}{\p y}}_{5} \\
  &- \underbrace{\averper{v\frac{\p u}{\p y}}\frac{\p^2 \averper{U}}{\p y^2}}_{6} -
  \underbrace{\averper{v\frac{\p w}{\p y}}\frac{\p^2 \averper{W}}{\p y^2}}_{7} -
  \underbrace{\averper{\frac{\p u_i}{\p x_k}\frac{\p u_j}{\p x_k}\frac{\p u_i}{\p x_j}}}_{8} -
  \underbrace{\frac{1}{2}\frac{\p}{\p y}\left(\averper{v\frac{\p u_i}{\p x_j}\frac{\p u_i}{\p x_j}}\right)}_{9} \\ 
  &- \underbrace{\averper{\frac{\p u_i}{\p x_j}\frac{\p^2 p}{\p x_j \p x_i}}}_{10} -
  \underbrace{\averper{\frac{\p^2 u_i}{\p x_j \p x_k}\frac{\p^2 u_i}{\p x_j \p x_k}}}_{11} +
  \underbrace{\frac{\p^2\epsilon}{\p y^2}}_{12}.
 \end{split}
\end{equation}
The order of magnitude of the terms arising in (\ref{eq:dissipation}) because of the wall motion can be estimated through an analysis similar to one for the enstrophy equation (\ref{eq:enstrophy}). The magnitude of term 3 in (\ref{eq:dissipation}) is found as follows.
\begin{equation}
 \mbox{Term} \ 3: \quad  \averper{\frac{\p w}{\p x_i}\frac{\p v}{\p x_i}}\frac{\p \averper{W}}{\p y}
 = \left(
   \underbrace{\averper{\frac{\p w}{\p x}\frac{\p v}{\p x}}}_{3a} 
 + \underbrace{\averper{\frac{\p w}{\p y}\frac{\p v}{\p y}}}_{3b} 
 + \underbrace{\averper{\frac{\p w}{\p z}\frac{\p v}{\p z}}}_{3c} 
 \right)\frac{\p \averper{W}}{\p y},
\end{equation}
\[
 3a \sim \frac{vw}{\lambda_x^2}, \
 3b \sim \frac{vw}{\delta^2} \sim \frac{uw}{\lambda_x\delta}, \
 3c \sim \frac{vw}{\lambda_z^2} \sim \frac{uv}{\lambda_z \lambda_x}, \ 
 \frac{\p \averper{W}}{\p y} \sim \frac{A}{\delta}.
\]
Term $3b$ is dominant, so that
\[
 \mbox{Term} \ {3}:\averper{\frac{\p w}{\p x_i}\frac{\p v}{\p x_i}}\frac{\p \averper{W}}{\p y} 
 = \mathcal{O}\left(\frac{uw A}{\delta^2\lambda_x}\right).
\]
The magnitude of term 5 in (\ref{eq:dissipation}) is estimated as follows.
\begin{equation}
\label{eq:term5}
 \mbox{Term} \ 5: \quad  \averper{\frac{\p u_i}{\p z}\frac{\p u_i}{\p y}}\frac{\p \averper{W}}{\p y}
 = \left(
   \underbrace{\averper{\frac{\p u}{\p z}\frac{\p u}{\p y}}}_{5a} 
 + \underbrace{\averper{\frac{\p v}{\p z}\frac{\p v}{\p y}}}_{5b} 
 + \underbrace{\averper{\frac{\p w}{\p z}\frac{\p w}{\p y}}}_{5c} 
 \right)\frac{\p \averper{W}}{\p y},
\end{equation}
\[
 5a \sim \frac{u^2}{\delta\lambda_z}, \
 5b \sim \frac{v^2}{\delta\lambda_z}, \
 5c \sim \frac{w^2}{\delta\lambda_z}.
\]
Term $5a$ is dominant. It follows that
\[
\mbox{Term} \ 5:\averper{\frac{\p u_i}{\p z}\frac{\p u_i}{\p y}}\frac{\p \averper{W}}{\p y} = \mathcal{O}\left(\frac{u^2 A}{\delta^2\lambda_z}\right).
\]
It is found that
\begin{equation}
\label{eq:term7}
\mbox{Term} \ 7: \quad  \averper{v\frac{\p w}{\p y}}\frac{\p^2 \averper{W}}{\p y^2} = \mathcal{O}\left(\frac{uw A}{\delta^2\lambda_x}\right)
\end{equation}
because
\[
\averper{v\frac{\p w}{\p y}} \sim \frac{vw}{\delta} \sim \frac{uw}{\lambda_x}, \
\frac{\p^2 \averper{W}}{\p y^2} \sim \frac{A}{\delta^2}.
\]
Term 5a is estimated to be the largest one amongst the terms in (\ref{eq:dissipation}) induced by the wall motion. This result confirms the analysis of the turbulent enstrophy, where term 3, of the same order of magnitude, emerges as dominant.

%-------------------------------
\section*{Acknowledgements}

Claudio Ottonelli would like to express his gratitude to the Department of Mechanical Engineering at King's College London for the hospitality from October 2009 until February 2010. Yosuke Hasegawa greatly acknowledges the support from the Japan Society for the Promotion of Science (JSPS) Postdoctoral Fellowship for Research Abroad. We thank Dr Andrea Ducci for the enlightening discussions on the enstrophy production in \S\ref{sec:physics}. We are also indebted to Drs J. Dusting, F. Martinelli, and G. Papadakis for their useful comments.

\bibliographystyle{jfm}
%\bibliography{../../mq}

\end{document}